# Hole Scavenging and Photo-Stimulated Recombination of Electron-Hole Pairs in Aqueous TiO$_2$ Nanoparticles [a)]

I. A. Shkrob[*] and M. C. Sauer, Jr.

*Chemistry Division, Argonne National Laboratory, Argonne, IL 60439*

**Abstract**

It is shown that 532 nm and 1064 nm laser photoexcitation of trapped electrons generated by 355 nm photolysis of aqueous titania (TiO$_2$) nanoparticles causes rapid photobleaching of their absorbance band in the visible and near IR. This photobleaching occurs within the duration of the laser pulse (3 ns FWHM); it is caused by photoinduced electron detrapping followed by rapid recombination of the resulting free electron and a trapped hole. The quantum yield for the electron photobleaching is ca. 0.28 for 532 nm and ca. 0.024 for 1064 nm photoexcitation. Complete separation of the spectral contributions from trapped electron and hole is demonstrated using glycerol as a selective hole scavenger. When glycerol is added to the solution, some light-absorbing holes are scavenged promptly within the duration of the 355 nm photoexcitation pulse, some are scavenged at a slower rate over the first 200 ns after the 355 nm pulse, and the rest are not scavenged, even at high concentration of the scavenger (> 10 vol %). A reaction with chemi- and physi- sorbed glycerol would account for the prompt and the slow hole decay, respectively. The implications of these results are discussed.

_________________________________________________________________________________

## 1. Introduction.

Semiconducting metal oxide nanoparticles, such as aqueous $TiO_2$ nanoparticles, find numerous applications in photovoltaics and photocatalysis [1,2,3]. Photoexcitation of these nanoparticles in the UV yields electron-hole pairs that rapidly recombine; this recombination competes with trapping of the free charges by coordination defects at the surface and by lattice defects in the nanoparticle bulk. Most of these electrons and holes recombine within the first few tens of picoseconds after the photoexcitation event. [4,5,6] However, conduction band (CB) electrons injected from dye molecules chemisorbed at the surface of nanocrystalline $TiO_2$ films can be observed for more than 1 ns, using time-resolved infrared (IR) [7] and terahertz (THz) [8] spectroscopies, suggesting relatively inefficient electron trapping rates in these high-quality films. Recent results of Turner et al. [8] suggest that these (nominally, "conduction band") electrons exhibit features that are more consistent with Anderson localization and transport by thermally-activated hopping. In particular, THz conductivity is dominated by backscattering, and the electrons have strikingly non-Drude behavior. "Hot" electrons observed in the first 330 fs after the electron injection are more Drude-like, [8] and only such species can be regarded as "free" CB electrons.

In $TiO_2$ nanoparticles, the primary charges rapidly descend into traps, eventually yielding well-localized species that decay slowly by bimolecular and geminate recombination on a time scale of $10^{-9}$-$10^{-1}$ s. [4,5,6,9-14] In the low charge density regime, the time profile of this decay exhibits dispersive $t^{-1/2}$ kinetics typical of trap-to-trap hopping observed for other disordered systems. [14,15] In the following, these electrons and holes will be referred to as "mobile" since these species can migrate, albeit slowly, and recombine with each other. We stress that these localized species are different from the *free* carriers. While some authors believe that *free* CB electrons exist in $TiO_2$ nanoparticles for indefinite time, time-resolved conductivity studies suggest the opposite (see below). With other techniques, it may be difficult to distinguish between the "free" carriers and band-tail charges in shallow traps (from which thermal emission of the free carriers can readily occur). The preceding does not mean that metastable CB electrons cannot occur in $TiO_2$ under favorable conditions (e.g., such species were observed in the diffuse reflectance IR spectra of heat- and oxygen- treated, UV-illuminated hydrated rutile surfaces); [16] the point is that such long-lived species do not occur in aqueous $TiO_2$ nanoparticles.

Following the rapid trapping of free carriers by traps in the $TiO_2$ nanoparticles (< 200 fs, [4] <330 fs [8]), the mobility of the electron is orders of magnitude lower [17,18] than the drift and Hall electron mobilities observed in high-quality anatase films. [19] From the extensive data on microwave (GHz) photoconductivity obtained by Warman and coworkers [17,18] and Martin et al. [20] it appears that the charge migration in these nanoparticles (typical mobilities are $5 \times 10^{-4}$ cm$^2$/Vs for holes and $10^{-3}$-$10^{-1}$ cm$^2$/Vs for electrons, depending on the sample) [17,18] is facilitated either by small polaron hopping or by thermal emission of free carriers from shallow traps; [18] direct trap-to-trap hopping is believed to be inefficient at room temperature (though, it may be more important at cryogenic temperatures). [18] The depth of the traps has been estimated by different authors as being from 60-100 meV [20,21] to 400-600 meV [16,22,23] (the higher estimates are clearly incompatible with the thermal emission mechanism mentioned above). Many types of electron traps may coexist in the aqueous nanoparticles, and the scatter in the estimates reflects the fact that different electron populations are observed in different experiments. No single species can explain all of the results.

Most of the dynamic and structural studies of "electrons" and "holes" in titania nanoparticles are done on these *trapped* species. Though the trapped charges participate in photocatalysis, they are less reactive than their rapidly migrating, energetic precursors. For this reason, photocatalytic reactions that involve molecules in the solution are relatively inefficient, [1,2] as only trapped charges can react with such molecules. On the other hand, photooxidation and photoreduction of molecules that are chemisorbed on the nanoparticle surface can be very rapid and efficient (see, for example, refs. [2,4,7,12]) as these processes are due to reaction with the free carriers.

The predominant trapped-electron centers are interior and surface $Ti^{III}$ centers that yield, respectively, narrow symmetric and broad asymmetric lines in Electron Paramagnetic Resonance (EPR) spectra from UV-irradiated frozen solutions of $TiO_2$ nanoparticles at 4-77 K. [3,24,25,26] In the absence of the hole scavenger, most of the electrons in $TiO_2$ nanoparticles recombine with the holes; the only species that survive in the low-temperature photoilluminated samples are interior electrons (ca. one center per nanoparticle). In the presence of hole scavengers, this recombination is inhibited and both types of $Ti^{III}$ centers are observed by EPR spectroscopy. [3,24,26] In acidic solutions, most of the electrons are trapped at the surface of $TiO_2$ nanoparticles. These surface centers were not observed in alkaline solutions [3,24] and in nanoparticles whose surface is modified by chelating agents, such as cysteine. [3,25] Frequently, the surface modification results in the formation of new types of surface electron centers, with different EPR spectra. [3,25,26]

The structure of the hole center(s) has not yet been definitively established. The holes are trapped mainly at the surface of the $TiO_2$ nanoparticle and, consequently, their EPR spectra depend strongly on the surface treatment. All of these centers seem to be $O$ $2p$ radicals, such as $Ti^{IV}-O^{\bullet}$ radical and adsorbed hydroxyl, $HO^{\bullet}_{ads}$ [3,24,26] According to Ishibashi et al., [27] the quantum yield of the latter species is relatively small (ca. 0.7% of the quantum yield for the iodide-oxidizing mobile hole, see below), however, this radical is more reactive towards certain substrates. Gao et al. [28] demonstrated that only $HO^{\bullet}_{ads}$ can oxidize aliphatic monohydroxy alcohols (such as methanol) in aqueous $TiO_2$ nanoparticle solutions, whereas the predominant type of the hole in photoexcited $TiO_2$ does not react with these alcohols on the microsecond time scale. [29] Interestingly, rapid reaction of holes with alcohols was observed for clean $TiO_2$ surfaces exposed to alcohols, both as liquid and vapor, by time resolved GHz conductivity [18] and IR spectroscopy, respectively. [30] These results, as well as EPR studies of Micic et al. [26] suggest that the holes are scavenged by alkoxy groups anchored at the titania surface [30,31]. Hydrolysis of these groups by water drastically reduces hole scavenging efficiency.

Surprisingly little has been done to ascertain light absorption properties of these electrons and holes since the initial transient absorbance (TA) studies carried out in the early 1980's. [6-12] At $pH$=3-4, the absorption spectrum of photoexcited $TiO_2$ nanoparticles is an asymmetric broad band with a maximum at 600-650 nm that extends to the IR. [6] In acidic or alkaline solutions containing hole scavenger (e.g., polyvinyl alcohol), [9-12] it is a broader, more symmetric band centered at 900-1000 nm. [32] A very similar band was observed by (i) electrochemical reduction of $TiO_2$ nanocrystalline films (at any $pH$ and salinity of the solution in contact with the film), [33] (ii) by injection of radiolytically-generated hydrated electrons into TiO$_2$ nanoparticles in *pH*=3 solutions, [34] and (iii) in steady-state UV photolysis of the solutions containing a hole scavenger (typically, an alcohol [9-13] or carboxylic acid [35]). From these convergent results, it appears that the 900 nm band is from the relatively long-lived trapped *electrons*. However, it remains unexplained why the TA spectra observed from photoexcited TiO$_2$ nanoparticles look different from this characteristic 900 nm spectrum.

Recently, TA spectra of photoexcited bare TiO$_2$ nanoparticles (in transmission) and nanocrystalline films (in diffuse reflection) in the 400-700 nm region became available on the pico- and femto- second time scales, by using pump-probe ultrafast laser spectroscopy. [4,5,36,37] Though no two such TA spectra look alike (as there is considerable variation with the particle size, surface treatment, etc.), all of these picosecond TA spectra exhibit considerable evolution over the first few tens of picoseconds and change little afterwards, suggesting that the descent to deeper traps and equilibration of the electrons and holes between different traps is largely over in 1-10 ps. The spectral features observed on nanosecond to millisecond time scales are fully developed by that delay time. [4,6]

Most workers, following the original suggestion by Graetzel and coworkers [6] and Henglein, [9] assumed that the 400-1000 nm absorbance from photoexcited TiO$_2$ nanoparticles originated from trapped *electrons* exclusively. Such a rationale implies that more than one species of light-absorbing trapped "electrons" (i.e., $Ti^{III}$ centers) contributes to the TA spectra: one kind of traps yields the 900 nm centered spectrum, another kind yields the 650 nm centered spectrum. Based on the EPR spectroscopy studies discussed above, [3,24,26] one may expect that the 900 nm band is from the electrons that are trapped in the nanoparticle interior, whereas the 650 nm band is composite (originating both from the interior and surface-trapped electrons). Such a rationale, however, seems to be at odds with the fact that the TA spectrum changes slightly, if at all, over the first 1 ms after the UV photoexcitation. If different kinds of trapped electrons originate from different environments, why would all such electrons decay in exactly the same way over many decades in time? If different subspecies of the "electron" originate from different traps on the same nanoparticle, what prevents the more energetic "electron" from descending to deeper traps?

5.

In this work, we demonstrate selective removal of a subset of light-absorbing species in acidic solutions of photoexcited $TiO_2$ nanoparticles by using glycerol as a scavenger. We present several lines of evidence suggesting that a trapped *hole* rather than the "electron" is the progenitor of the 650 nm band. The absorption spectrum of the hole in the visible resembles the spectrum of oxidizing species in "platinized" $TiO_2$ nanoparticles observed by Bahnemann et al. [10,11,12] Our study suggests that the (trapped) electron in photoexcited $TiO_2$ nanoparticles has the same spectrum as the electron obtained by radiolytic (e.g., ref. 34) and electrochemical (e.g., ref. 33) reduction of $TiO_2$ nanoparticles and nanocrystalline films. Photoexcitation of this electron at 532 and 1064 nm causes its detrapping followed by rapid recombination with the hole. Scavenging of the hole by glycerol reduces the quantum yield of this photostimulated recombination to the same degree it reduces the absorption of the hole. To save space, Figs. 1S to 7S are placed in the Supporting Information.

## 2. Experimental.

Transient absorbance was observed following 355 nm laser photoexcitation of $N_2$-saturated $2.4 \times 10^{-4}$ M aqueous solution of 4.6±0.5 nm diameter anatase nanoparticles (1400 units of $TiO_2$ per particle) at *pH*=4. This solution was prepared and characterized as described in ref. [38]. The sample contained 45 ppm of aluminum by volume (that would be equivalent to 7 atoms per particle if $Al^{III}$ were incorporated into the $TiO_2$ nanoparticle) and it is possible that vis-absorbing holes observed in this study originated from a $Ti^{IV} - O - Al^{III}$ or some other impurity center (section 4.3). All other chemicals were obtained in their purest form available from Aldrich and used as received.

The excitation pulse L1 (355 nm, 3 ns FWHM, 8 mJ ) that was used to induce charge separation in $TiO_2$ was derived from the third harmonic of a Nd:YAG laser (Quantel Brilliant). The second excitation pulse (L2) of 532 nm (6 ns FWHM, 10-50 mJ) and 1064 nm light (6 ns FWHM, 90 mJ) used to photoexcite trapped electrons in $TiO_2$ was derived from the second and the first harmonic, respectively, of another Nd:YAG laser (Continuum model 8010). This electron-excitation pulse was delayed in time relative to the 355 nm pulse; the time jitter between these pulses was < 3 ns.

The analyzing light from a superpulsed 70 W Xe arc lamp was filtered and then crossed at 45° with the 355 nm laser beam inside a 1.35 mm optical path flow cell with suprasil windows (a 5.6 mm diameter aperture was used for light masking). The duration of the analyzing light pulse was 300 μs. The optical density (OD) of the solution at 355 nm was 0.8. The 532 nm (or 1064 nm) beam was crossed with the 335 nm beam at 25°, so that the analyzing light, the UV beam and the 532 (1064) nm beams made the angles of 31°, -14°, and +11° with the normal to the cell window, respectively.

The transmitted light was passed through an appropriate glass cutoff filter placed before the sample, focussed, passed through the sample, and then passed through another appropriate cutoff glass filter and a grating monochromator (SPEX Minimate; 450 to 950 nm). For detection at wavelengths $\lambda > 1$ μm, the analyzing light was passed through a 4 cm optical path water filter or a 400-700 nm bandpass dielectric filter to reduce the heat transfer to the sample. For detection at 1.0-1.35 μm, a set of narrowband (10 nm FWHM) interference filters at increments of 50 nm was used. The cutoff of 1.35 μm was due to the strong IR absorbance of water. For detection near 1 μm, a 0° dielectric mirror reflecting at 1064 nm was inserted before the detector to reduce the leakage of the first Nd:YAG harmonic. A fast Si photodiode (EG&G model FND-100Q) biased at -90 V was used to detect the TA for $\lambda \leq 1.1$ μm. For detection at $\lambda \geq 1.1$ μm, a fast Ge photodiode (Germanium Power model GMP-566) biased at -10 V or an InGaAs photodiode (Germanium Power model GAP-520) biased at -5 V was used. The latter photodiode is fast, but its temporal response is strongly wavelength dependent. [39] The Ge photodiode has a spectrally flat temporal response, [39] however, it is slower than the other two photodiodes, and TA kinetics acquired at delay time $t < 30$ ns were distorted. Furthermore, a thermal lensing effect induced by the absorption of 355 nm pulse unevenly refracted the IR light across the sample causing a parasitic transient that lived for the first 20 ns after the UV photoexcitation. Thus, our near IR spectra were not trustworthy at these short delay times.

The photodiode signal was amplified 10 times using a Comlinear model CLC449 1.2 GHz opamp and sampled using a Tektronix TD360 digital oscilloscope or a Tektronix DSA-601 digital signal analyzer. Absorption signals as small as $10^{-4}$ can be studied with this setup. Light emission from the sample induced by the 355, 532, and 1064 nm laser light alone was subtracted from the absorbance signal. In the following, 355 nm light induced TA kinetics are denoted $\Delta OD_\lambda(t)$, transient absorbance induced by both 355 nm (L1) and 532 nm or 1064 nm (L2) light is denoted $\Delta OD_\lambda^{L2}(t_{21};t)$, where $t_{21}$ is the delay time between the laser pulses L2 and L1, and

$$\Delta\Delta OD_\lambda(t_{21};t) = \Delta OD_\lambda^{L2}(t_{21};t) - \Delta OD_\lambda(t) \qquad (1)$$

is the difference trace.

The two-pulse experiment was complicated by laser-induced precipitation of $TiO_2$ nanoparticles: Regardless of the flow rate of the solution, prolonged 355 nm laser irradiation caused the deposition of a thin, transparent film of $TiO_2$ particles on the cell windows, due to the action of a shock wave generated by the 355 nm laser. This deposit can be removed by boiling the suprasil windows in concentrated $HNO_3$ for 2-3 *h*. Though this deposit had no discernible effect on the TA kinetics observed after the 355 nm pulse, a strong, persistent TA signal in the visible was observed when this deposit was photoexcited by 532 nm light 10-100 ns after the 355 nm pulse. For this reason, the exposure of the cell windows was limited to < 1000 pulses and the cell was periodically filled with pure water to monitor for the formation of the film deposit. For the same reason, a cell with detachable optical windows has been used. The typical flow rate was 0.5-2 $cm^3$/min and the repetition rate of the laser was 1-2 Hz.

In some experiments, a hole scavenger (1-5 vol % glycerol) was added to the reaction mixture. The C-centered radicals generated in the photooxidation of glycerol by holes in the $TiO_2$ nanoparticles do not absorb light at $\lambda$>400 nm. To prevent the buildup of a permanent trapped-electron absorbance during continuous laser photolysis, the sample was saturated with $O_2$. The oxygen reacts with the electrons on a millisecond time scale, [40] making the photosystem reversible. Control experiments with $N_2$-saturated $TiO_2$ nanoparticle solutions showed that the oxygen does not react with the electrons on the time scale of our study (< 10 μs, Figs. 1S and 2S). In another control experiment, an $N_2$-saturated, glycerol-containing solution of $TiO_2$ nanoparticles was continuously illuminated with 300-400 nm light from the Xe arc lamp for 1 *h* to induce a strong absorbance from the trapped electrons (the *OD* of this solution at 400-900 nm was 0.07-0.1) and then TA kinetics were obtained following the 532 nm laser photoexcitation. No 532 nm light induced transient photobleaching was observed, nor did we observe any photobleaching when the UV-photolysed $N_2$-saturated glycerol solution was illuminated with 1-5 mW of visible light for 1-2 *h*.

Since under the conditions of our TA experiment < 1-2 electron-hole pairs per particle were present at the end of the 355 nm pulse, [31] scavenging of the hole by glycerol results in a long lived TA signal from electrons which lack hole partners on the same nanoparticle (referred to as the "persistent" electron in the following).

**3. Results.**

*3.1. Single pulse 355 nm photoexcitation.*

Fig. 1 shows typical TA kinetics observed in 355 nm laser photolysis of aqueous $TiO_2$ nanoparticles ($\lambda = 700$ nm). The initial "spike" whose time profile closely follows the temporal response function of the setup (Fig. 1a) is rapidly succeeded by the power law $t^{-\alpha}$ decay with $\alpha \approx 0.46$ (Fig. 1b). This slow decay kinetics can be followed out to at least 100 μs. Exactly the same time profiles of transient absorbance are observed across entire the visible, for $t > 5$ ns (Fig. 1S), and in the near IR, for $t > 20$ ns (Fig. 2S). Since all of these time profiles are identical, there is no time evolution of the TA spectrum. In Fig. 2, normalized TA spectra in the visible obtained at different delay times are plotted together (Fig. 3a shows the typical evolution of the TA spectra prior to the normalization). All of these spectra are identical within the experimental scatter (see also Fig. 4a below). As seen from Fig. 2S, the same applies to the TA spectra obtained in the near IR (*filled diamonds* in Fig. 2). For $\lambda > 900$ nm, minor changes in the spectrum were observed within the first 20 ns. These changes turned out to be a combined effect of the slow photoresponse of the Ge photodiode [39] and wavelength-dependent refraction of the analyzing light by a thermal lens in the sample (section 2).

Qualitatively similar TA spectra (400-900 nm) and TA kinetics (< 1 μs) were observed for $TiO_2$ nanoparticles in acidic solutions by previous workers. [9-12] Unlike some of these workers (see, for example, Rothenberger et al. [6] and Bahneman et al.

[12]) we found no evidence of time evolution for TA spectra over the first 10 μs after the photoexcitation, both in $N_2$- and $O_2$-saturated solutions. Neither did we observe spectral changes when electron scavengers, such as dimethyl sulfoxide, tetranitromethane, $SF_6$, and $N_2O$ were added to the aqueous solution. Apparently, trapped electrons in the $TiO_2$ nanoparticles reacted too slowly with these scavengers. Yet the reaction eventually occurred since addition of these scavengers prevented the development of permanent blue coloration from the trapped electron in the presence of hydroxylic hole scavengers. We stress that the signal-to-noise ratio in our TA kinetic traces was better than in the previous experiments, and we are confident that if there were a spectral evolution similar to that observed in refs. 6 and 12, it would be observed using our TA setup.

As mentioned above, most researchers attribute the spectrum shown in Fig. 2, trace (iii), to trapped electrons, though some studies suggested that trapped holes may also absorb in the visible. [10,11,12] Our results suggest that the TA spectrum shown in Fig. 2 is a composite spectrum of the absorbances from the electron and the hole. The resulting TA spectrum can be decomposed into the individual components by addition of a suitable hole scavenger, as discussed below:

Addition of 5 vol % of hole scavenger, glycerol, to oxygenated solution of $TiO_2$ nanoparticles (see section 2) strongly reduces the TA signal in the visible and simultaneously increases the TA signal in the near IR, as shown in Fig. 3b (compare with the TA spectra in Fig. 3a obtained under the same excitation conditions and time windows for an $O_2$-saturated solution without the hole scavenger). Even the prompt TA spectrum obtained at the end of the 3 ns FWHM, 355 nm pulse is quite different from the TA spectrum shown in Fig. 3a. Further evolution occurs within the first 200 ns after the 355 nm photoexcitation pulse; at later delay times the TA spectrum stops changing (Fig. 3b). Importantly, the addition of the hole scavenger has almost no effect on the *shape* of the TA spectrum for $\lambda$>900 nm. This point can be demonstrated by comparing the TA spectra obtained at different delay times at 900-1300 nm, as shown in Fig. 3S. In Figs. 4a and 4b, the TA spectra shown in Fig. 3 were normalized at 900 nm. Without glycerol, all normalized the spectra are the same. For the glycerol solution, a spectral component in the visible is missing from the onset. At later delay times, more of this component is missing, as the species which absorbs light in the visible decays on a sub-microsecond time scale. This point is illustrated in Fig. 4S, where 450-950 nm TA kinetics in 5 vol % glycerol solution are normalized at 200 ns (by which delay time the spectral evolution is over). These normalized kinetics diverge in the first 50 ns after the photoexcitation, suggesting rapid spectral evolution on this short time scale.

To obtain the spectrum of the short-lived species, normalized TA spectra obtained at $t<200$ ns (Fig. 4b) were subtracted from the normalized spectrum obtained at $t>200$ ns (Fig. 5a, trace (ii)). As shown in Fig. 5b, the resulting difference traces do not change in shape with the delay time, which implies that the missing spectral component (trace (i) in Fig. 5b) originates from a *single* light-absorbing species. Eventually (at $t>200$ ns), all of this species decays and the spectrum shown by trace (ii) in Figs. 5a and 5b is obtained. This "final" TA spectrum does not depend on the glycerol concentration (1 to 10 vol %). Since the carbon-centered radicals derived from glycerol do not absorb in the visible, this final spectrum is mainly from a trapped electron (though, as shown below, a small fraction of holes also contributes to the long-lived TA signal).

Spectra from trapped electrons similar to the TA spectrum shown by trace (ii) in Fig. 5b have been observed by other researchers, in particular, (i) in laser photolysis of basic solutions of $TiO_2$ nanoparticles, [32] (ii) in steady-state UV photolysis (e.g., refs. 9, 32, and 35) and pulse radiolysis [34] of acidic $TiO_2$ solutions, and (iii) by spectrophotometry of electrochemically reduced thin $TiO_2$ films in contact with basic or acidic solutions [33]. In all of these experiments, the trapped holes either were not generated or promptly scavenged after their generation. Apparently, it is the presence of the hole on the aqueous $TiO_2$ nanoparticle which is responsible for the difference in the TA spectra shown in Figs. 3a and 3b. The difference traces shown in Fig. 5a thereby give the absorption spectrum of the hole (trace (i) in Figs. 2 and 5b). These spectra, with peak absorbances at ca. 480 nm and gradual decrease in absorbance towards the near IR, strikingly resemble the absorption spectra of the species obtained by Bahnemann et al. [10,11,12] by photoexcitation of "platinized" $TiO_2$ nanoparticles (section 4.3). The islands of metallic Pt deposited on the nanoparticle surface serve as efficient electron traps, and Bahnemann et al. attribute the TA spectra observed from the "platinized" $TiO_2$ nanoparticles to holes.

Importantly, a weighted sum of the spectral contributions from the "hole" (trace (i) in Fig. 5b) and the "electron" (trace (ii) in Fig. 5b) shown in Figs. 2 and 5a *exactly* reproduces the TA spectrum observed in the $TiO_2$ nanoparticle solution that does not contain glycerol (trace (i) in Fig. 5a and trace (iii) in Fig. 2). Thus, both of the spectral components are also present in the solution that does not contain glycerol.

The data shown in Fig. 4b can be used to estimate the proportions of the hole scavenging during and after the photoexcitation pulse. The normalization at 900 nm, where the hole does not absorb, allows a determination of the decrease in the absorption in the visible caused by scavenging of the hole. The absorbances for the longest time window shown in Fig. 4b are essentially all due to electrons only, because as shown below, only 5-10% of the holes are not scavenged by 5 vol % glycerol. Therefore, the difference between traces (i) and (ii) is a measure of the "total" scavengeable holes. For example, the difference between trace (i) and the 9-13 ns data *(open circles)* is a measure of the holes promptly scavenged during the 355 nm laser pulse. In this way we obtain that ca. 45-50% of the light absorbing holes are scavenged within the duration of the 355 nm photoexcitation pulse, while another 45% are scavenged over the first 200 ns after the photoexcitation (these percentages are given relative to the electron concentration present at the beginning of this reaction, Fig. 4b). Since most of the holes present by the end of the 355 nm pulse decay by recombination rather than this slow scavenging (see Fig. 9 below), such a reaction contributes very little to the net yield of persistent electrons at $t > 200$ ns.

As the scavenger concentration increases (Figs. 6a and 5S), the fraction $\Phi_e$ of trapped electrons observed at $\lambda \geq 900$ nm that avoid recombination with the hole and persist at $t > 300$ ns steadily increases from ca. 10% towards ca. 50% (see Figs. 7a and 5S(a) and Fig. 2 in ref. 31). This fraction was determined by taking the ratio of the near IR absorbance at 300-350 ns and the (maximum) prompt absorbance $\Delta OD_\lambda^{max}$ at the end of the 355 nm pulse. Exactly the same fractions $\Phi_e$ were obtained at $\lambda$=900 nm, 1100

12.

nm, and 1300 nm (Fig. 5S(c)), i.e., across the whole spectral range where the TA spectrum is dominated by the electron absorption. The increase in the fraction $\Phi_e$ of persistent electrons follows the Stern-Volmer law (Figs. 7a and 5S(c))

$$\Phi_e \approx A + B/(1 + K_b[glycerol]) \tag{2}$$

with the half-effect concentration $K_b^{-1}$ of 0.435 M ($A$ and $B$ are empirical constants). Very similar behavior was observed for other polyhydroxy hole scavengers, such as carbohydrates. [31] As shown elsewhere, the constant $K_b$ varies over several orders of magnitude, depending on the number of the anchoring $HO$ groups. On the other hand, addition of 1-50 vol % of monohydroxy alcohols has no effect on the TA kinetics observed in the visible (where both the hole and the electron absorb light) or near IR (where only electrons absorb light). This result has been previously obtained by Rabani et al., [29] and it is fully confirmed in the present work. The same negative result was obtained for acetic and malonic acid, which were commonly used as hole scavengers in previous photochemical studies. [35] Neither the TA signal from the hole decreases, nor the TA signal from the electrons increases. Apparently, hole scavenging by these compounds is slow and inefficient.

Importantly, when the plateau absorbance attained at $t > 300$ ns is subtracted from the 900-1300 nm kinetics and the resulting difference traces are normalized at the TA signal maximum, the resulting $f_\lambda(t)$ kinetics

$$f_\lambda(t) = \frac{\Delta OD_\lambda(t) - \Delta OD_\lambda(t = 330ns)}{\Delta OD_\lambda^{max} - \Delta OD_\lambda(t = 330ns)} \tag{3}$$

are identical within the experimental error. This point is illustrated in Fig. 7b (for $\lambda = 900$ nm) and Figs. 5S(a) and 5S(b) (for $\lambda = 1100$ nm and $\lambda = 1300$ nm, respectively). We conclude that the decay kinetics of electrons in the presence of a hole scavenger are a weighted sum of a constant and the decay kinetics obtained in the absence of the hole scavenger.

These observations suggests that there is a subset of holes which are scavenged very rapidly by the glycerol (well within the duration of the photoexcitation pulse) or,

13.

perhaps, whose generation is prevented by the presence of glycerol (if the latter reacts with their short-lived precursor). [41] Thus, in the glycerol solution, the TA signal at $\lambda \geq 900$ nm originates from two types of trapped electrons: (i) the persistent electron residing on a nanoparticle with a hole removed or transformed in a reaction with the scavenger and (ii) relatively short-lived electron residing on the nanoparticle which still has a hole. Naturally, for the latter electrons, the decay kinetics are exactly the same as those for the electron on a $TiO_2$ nanoparticle in the aqueous solution.

The same effect, *viz.* the relative increase in the long-lived, persistent TA signal at $t>200$ ns with the addition of a hole scavenger, was observed across the entire visible range (where both the electron and the hole absorb light); see Figs. 6b, 6c, and 7a. The magnitude of this effect increases with increasing wavelength $\lambda$ of the analyzing light. In Fig. 7a, the concentration dependencies for $\Delta OD_\lambda$ at 330-350 ns for $\lambda = 470$, 600, and 900 nm are fit by the dependencies given by the left hand side of eq. (2) with the same constant $K_b$ for all three wavelengths. This plot indicates that the reaction with glycerol trades the spectral contribution from the hole for the spectral contribution from the persistent electron which, once formed, does not decay on the microsecond time scale. The same point is demonstrated by Fig. 8a, where the "final" absorbances $\Delta OD_{470}$ and $\Delta OD_{600}$ attained at $t = 330\text{-}350$ ns are plotted as a function of the "final" optical density $\Delta OD_{900}$ for several glycerol concentrations. These plots are linear. The intercepts of these plots correspond to the 470 nm and 600 nm absorbances that originate from the holes that cannot be scavenged by glycerol (i.e., whose absorbances cannot be traded for the electron absorbance) even when a large concentration of glycerol is added to the reaction mixture. The TA signals from these "unscavengeable" holes are shown in Fig. 8b (with the arrows pointing to the corresponding values). Traces (i) and (ii) in Fig. 8b are normalized spectra of the "electron" and the hole given by traces (i) and (ii) in Fig. 2, respectively. Apparently, this "electron" spectrum is, in fact, composite, because some light-absorbing holes are not scavenged by glycerol on the microsecond time scale. The "scavengeable" (> 90 %) and "unscavengeable" (< 10%) holes have similar TA spectra. Assuming that these absorption spectra are identical, the residual hole spectrum trace (ii) can be subtracted from the "final" spectrum, trace (i), in the same figure. This procedure gives a bell-shaped spectrum of the trapped electron alone (trace (iii) in Fig. 8b). The same spectrum is shown in Fig. 2, trace (iv). Since any TA spectrum which is a linear combination of traces (i) and (ii) in Fig. 2 is also a linear combination of traces (ii) and (iv) in the same figure, our conclusion that the TA spectrum in water (trace (iii) in Fig. 2) is the composite spectrum of the electron and the hole is still correct.

As mentioned above, ca. 45-50% of the vis-light absorbing holes are scavenged by glycerol promptly and the remaining holes are scavenged slowly over the first 200 ns after the photoexcitation (Fig. 4b). The kinetics of this slow hole-scavenging reaction can be observed by normalizing the kinetic trace obtained in the visible (where both the electron and the hole absorb, Fig. 6b) and at $\lambda=900$ nm (where only the electron absorbs, Fig. 6a) by the "final" absorbance attained at $t=330$-$350$ ns and then taking the difference of these two normalized traces:

$$g_\lambda(t) = \frac{\Delta OD_\lambda(t)}{\Delta OD_\lambda(t=330\ ns)} - \frac{\Delta OD_{900}(t)}{\Delta OD_{900}(t=330\ ns)} \quad (4)$$

This procedure removes the contribution to TA from the electrons and "unscavengeable" holes, i.e. the resulting traces are the decay kinetics of the "scavengeable" holes. The resulting traces are shown in Fig. 9. To facilitate the comparison between these kinetics, the traces were normalized at the maximum. It is remarkable that these decay kinetics barely change as a function of glycerol concentration. Furthermore, most of this decay is due to recombination of the holes with the electrons. The recombination decay of these electrons in the absence of scavenger (as observed at $\lambda=900$ nm) is given by trace (i) in Fig. 9; the decay kinetics of the holes (at $\lambda=600$ nm) can be obtained by weighting these kinetics by an exponential function with a time constant $\tau_h$ of ca. 130 ns (trace (ii) in Fig. 9).

The fact that this "slow" hole-scavenging reaction shows almost no dependence on the glycerol concentration suggests that this reaction does not involve a glycerol molecule in solution, as otherwise the kinetic mass law would be obeyed. Rather, it should involve a *physisorbed* glycerol molecule. Once the coverage of the nanoparticle by glycerol is total (which occurs at 0.3-0.7 vol %), the reaction rate cannot increase further. Thus, we suggest that hole scavenging by glycerol (and polyols, in general) is two-stage: [31] First, the holes are promptly scavenged by *chemisorbed* molecules that chelate the titanium cation at the nanoparticle surface (see ref. 31 for more detail). This reaction is very rapid ( taking perhaps < 10 ps); [4] it competes with charge recombination and trapping of the hole by other surface defects. Once this ultrafast reaction is over, the surviving holes (which descended to deep traps) slowly react with physisorbed polyol molecules by H-abstraction or charge transfer. The residual holes (which are a small fraction of the total) are trapped by lattice defects in the subsurface or the interior of the nanoparticle and cannot react with physi- or chemi-sorbed polyols (on the microsecond time scale). Previously, Bahnemann et al. [12] used nanosecond flash photolysis to study hole scavenging by thiocyanate and dichloroacetate, and reached similar conclusions: Some holes react with the adsorbed hole scavengers promptly within the duration of the laser pulse (ultrafast pump-probe spectroscopy studies of Bowman et al. [4] suggest that for thiocyanate, this prompt reaction occurs on the femtosecond time scale), some holes react more slowly (at least, for dichloroacetate), and some holes do not react at all, at any concentration of the hole scavenger. While not all of our observations are in perfect agreement with the study of Bahnemann et al. [12] (e.g., we did not observe the effect of oxygen on the evolution of TA spectra), the general features of the hole dynamics observed in the present work (two-stage hole scavenging, different subspecies of these holes, the absorbance from the holes in the visible) are strikingly similar to those observed in ref. 12, despite the different experimental approach and the different nature of the hole scavengers.

Since the TA spectrum shown in trace (iii) in Fig. 2 is composite, an explanation is due as to why in the absence of a hole scavenger both of the species that contribute to this spectrum decay in exactly the same manner over many decades in time. This behavior is readily accounted for provided that the contributing absorbances originate from the electron and the hole, as suggested above. Indeed, if the latter species decay by recombination, there would always be a parity between their concentrations. Another piece of evidence favoring this scenario is given in the next section which deals with photostimulated recombination of the electron-hole pairs.

### *3.2. 532 and 1064 nm light induced photobleaching of the TA absorbance.*

Fig. 10a demonstrates a change in the TA kinetics induced by a short pulse of 532 light applied at delay time $t_{21}$=24 ns following 355 nm photoexcitation ($\lambda$ = 600 nm). In this plot, trace (i) shows the decay kinetics without the 532 nm pulse, trace (ii) shows the decay kinetics with the 532 nm pulse, and trace (iii) shows the difference kinetics given by eq. (1). As seen from Fig. 10a, the TA signal (trace (ii)) first rapidly decreases within the duration of this 532 nm pulse, then recovers on the same time scale as the $\Delta OD_{600}$ kinetics decays. Fig. 10a, trace (iv) shows the normalized difference trace (iii) (the recovery kinetics of the $\Delta\Delta OD_{600}$ signal) juxtaposed on the TA kinetics induced by the absorbance of the 355 nm light alone (trace (i)). As seen from Fig. 10a, these two kinetics are identical within experimental error, and this situation persists out to at least $t$=15 μs.

These observations suggest that the 532 nm photoexcitation promptly depletes at least one of the species that absorb 600 nm light within the duration of the 532 nm excitation pulse. However, not all of these species are photobleached; some remain in the reaction mixture well after the 532 nm photoexcitation and these remaining species have exactly the same decay kinetics as the 600 nm light absorbing species following 355 nm photoexcitation. This conclusion pertains not only to the TA kinetics observed at $\lambda$ = 600 nm, but to all TA kinetics observed between 400 nm and 1350 nm. Furthermore, it holds for both 532 nm and 1064 nm photoexcitation (Fig. 6S).

Let us introduce the fraction $q_\lambda$ of the TA signal that is promptly photobleached by the 532 nm (or 1064 nm) light, defined as the ratio

$$q_\lambda = -\Delta\Delta OD_\lambda(t_{21}; t \approx t_{21}) / \Delta OD_\lambda(t \approx t_{21}) \qquad (5)$$

of the photobleached TA signal to the TA signal at $t = t_{21}$. As explained above, this quantity is fully sufficient to characterize the entire photobleaching dynamics because the temporal behaviors of the $\Delta\Delta OD_\lambda(t_{12}; t)$ and $\Delta OD_\lambda(t)$ kinetics are exactly the same (traces (iii) and (i) in Fig. 10a, respectively). Due to this similarity, it is possible to fit both of these kinetics using the same smooth function (for instance, a biexponential dependence) and extrapolate these dependencies to $t = t_{21}$, thereby increasing the accuracy with which the quantity $q_\lambda$ is estimated. The resulting photobleaching fractions $q_\lambda$ obey the following simple rules, for both 532 nm and 1064 nm photoexcitation: For a given photon fluence of laser L2, the fraction $q_\lambda$ does not depend on the observation wavelength $\lambda$ (Fig. 10b) or on the delay time $t_{21}$ between the laser pulses L2 and L1. This fraction linearly increases with the photon fluence of laser L2. Addition of a hole scavenger, such as glycerol, (Fig. 11) causes a decrease in $q_\lambda$; this decrease becomes greater at longer delay times, $t_{21}$, between the laser pulses.

The linear increase in the photobleaching efficiency with increasing 532 nm (or 1064 nm) photon fluence indicates that the photoexcitation is 1-photon. The independence of the ratio $q_\lambda$ on the delay time $t_{21}$ means that a constant fraction of the 532 nm (or 1064 nm) light absorbing species is photoexcited at any time on the TA vs. time curve. The lack of wavelength dependence means that the profile of the photobleaching spectrum is identical to that of the species before the 532 nm or 1064 nm photoexcitation. That would not be remarkable if the TA spectrum shown in Fig. 2, trace (iii) originated from a single light-absorbing species depleted by the 532 nm or 1064 nm photoexcitation. However, as shown in the previous section, this does not seem to be the case. Thus, to account for our observations it is necessary to postulate that the 532 nm or 1064 nm photoexcitation of *one* of the light absorbing species depletes *both* of these species, in exactly the same ratio with which these two species were present prior to the photoexcitation. For example, as explained in section 3.1, the TA signal at 1064 nm is from trapped electrons only. If only electrons were depleted in the course of 1064 nm photoexcitation, the fractions $q_\lambda$ at $\lambda = 600$ nm (where both the electron and the hole absorb), and at $\lambda = 900$ nm (where only the electron absorbs), would be different. Experimentally, these two fractions are the same within 10%. The same argument pertains to the 532 nm photoexcitation. Thus, either the spectrum shown in Fig. 1, trace (iii) is from a single species (section 4.2), contrary to the arguments presented above, or exactly the same fraction of electrons and holes decays following 532 nm or 1064 nm photoexcitation. In the latter case, only photostimulated recombination can account for the observed behavior.

The effect of the hole scavenger on the photobleaching strengthens the latter explanation. A steady state concentration of trapped electrons in TiO$_2$ nanoparticles can be generated by continuous UV illumination of N$_2$-saturated aqueous solution containing a hole scavenger (in our case, glycerol). After 30 min illumination by 300-400 nm light, an optical density of 0.2-1 at 532 nm from trapped electrons was obtained (in a 1 mm optical path cell). When this solution was photolyzed using 532 and 1064 nm pulses, no transient photobleaching, either temporary or permanent, was observed. This negative result suggests that in the absence of a hole localized on the same TiO$_2$ nanoparticle, laser excitation of trapped electron does not lead to photobleaching of its absorbance. Rather, the electrons rapidly relax to the ground state, with no concomitant change in the absorbance.

The same point may be demonstrated differently, by carrying out a two pulse experiment on O$_2$-saturated solution containing 2-5 vol % of glycerol (O$_2$ was used for photoreversibility, see section 2). As shown in section 3.1, addition of 5 vol % glycerol causes ca. 45% decrease in the prompt TA signal from the trapped hole, due to rapid reaction with chemisorbed glycerol. Following the same trend, the ratio $q_\lambda$ decreases by ca. 50% at $t_{21}$=25 ns (Fig. 11). At later delay times, there is a slow decay of the TA signal from the holes due to their reaction with glycerol (Figs. 4b, 5b, and 9). The fraction $q_\lambda$ plotted as a function of the delay time $t_{21}$ tracks the hole scavenging (Fig. 11). We conclude that the presence of a hole on the same TiO$_2$ nanoparticle is required for the decay of the photoexcited electron; in the absence of the hole, there is no decay. Conversely, any reaction that depletes the hole reduces the photobleaching efficiency.

We turn now to the quantum yield for the 532 and 1064 nm photobleaching. This quantity can be calculated from the known photobleaching fraction $q_\lambda$ and the laser fluence provided that the molar absorptivity of the species at the photoexcitation wavelength is known. Unfortunately, there is a large scatter in estimates for the molar absorptivity of trapped electrons in aqueous TiO$_2$ nanoparticles. Photochemical studies by Graetzel and co-workers suggested $\varepsilon_{620}$=1200 M$^{-1}$ cm$^{-1}$ in acidic [6] and $\varepsilon_{780}$=800 M$^{-1}$ cm$^{-1}$ in alkaline solutions. [32] Pulse radiolysis measurements of Safrany et al. [34]

suggested $\varepsilon_{800}$=700 M$^{-1}$ cm$^{-1}$. Spectroelectrochemical measurements of Fitzmaurice and coworkers gave $\varepsilon_{700}$=900-1000 M$^{-1}$ cm$^{-1}$. [33] Using the absorption spectrum of the electron given by trace (iv) in Fig. 2 and the estimate given by Safrany et al., [34] the extinction coefficients for the electron at both 532 nm and 1064 nm are ca. 480 M$^{-1}$ cm$^{-1}$, and the quantum yields for photobleaching are ca. 0.28 and 0.024, respectively.

## 4. Discussion.

*4.1. Synopsis.*

Here we briefly summarize our observations. The TA spectrum in the visible is composite. Two species contribute to this spectrum: (i) trapped electron, which has roughly the same absorption spectrum as the species generated by electron injection and reduction of TiO$_2$ nanoparticles and films (e.g., refs. 32, 33, and 34) and (ii) trapped hole, which has an absorption spectrum in the visible. [10,11,12]. These two species decay by recombination that proceeds over many decades in time; the resulting spectrum does not evolve in time because both species simultaneously decay in the recombination event. When the electron is detrapped by 1064 nm or 532 nm photoexcitation, the resulting free, mobile electron rapidly recombines with the hole, and both species disappear from the TA spectrum. The quantum yield of the photobleaching by visible light is large, ca. 0.28 for 532 nm photoexcitation. When the hole is scavenged, photostimulated recombination is suppressed.

Recently, Boschloo and co-workers [23] observed longer time of flight and decreased transferred charge when the photocurrent induced by 308 nm laser excitation of TiO$_2$ films was measured under *continuous-wave* irradiation by light with $\lambda > 420$ nm. The authors explained this behavior by "emptying of the initially filled traps by the visible light". [23] As suggested by our results, this "trap emptying" is followed by electron-hole recombination that depletes the charge flowing through the conductivity cell.

The light-absorbing hole can be selectively depleted by reaction with chemi- and physi- sorbed polyols, [9,10,11,31] such as glycerol. Most of this hole scavenging is very

20.

rapid (occurring in < 3 ns), and there are good reasons to believe that it involves a chemisorbed polyol molecule at the nanoparticle surface (see ref. 31 for more discussion). However, this scavenging is not 100% efficient, even at high polyol concentration (> 5 vol % of glycerol). Ca. 45% of the holes survive this rapid process by descent into deep traps at the nanoparticle surface and slowly react with physisorbed glycerol molecules over the first 200 ns after the photoexcitation. There is also a small subset of light-absorbing holes (<10% of the total population) that are not scavenged by glycerol in either reaction; these are holes trapped in the nanoparticle interior (or, perhaps, the subsurface). Given the low concentration of electron-hole pairs generated in our experiments (1-2 pairs per nanoparticle), the depletion of the hole by a reaction with glycerol leaves an electron which does not have a recombination partner on the same nanoparticle. Such electrons persist in solution for many tens of microseconds.

While these scenarios are consistent with our observations and find support in other studies, there is a troubling aspect to such an interpretation of the results. Namely, it is commonly believed that trapped holes in titania nanoparticles do not absorb light in the visible. Most researchers, following the original study by Graetzel and coworkers, [6] assume that only *electrons* absorb in the visible, and this viewpoint has been supported by pulse radiolysis studies [42,43] in which the holes were shown to absorb at $\lambda < 400$ nm (section 4.3). Although there have been several studies the results of which are contrary to the assumption that only electrons absorb in the visible, [10,11,12,36] it is prudent to consider alternative scenarios in which the hole does not absorb in the visible to explain our results.

### *4.2. The alternative scenarios: Two-electron models.*

In this section we examine several scenarios in which the spectral transformations observed in section 3.1 are interpreted in terms of two or more light-absorbing electron species. In these scenarios, the hole is assumed to absorb in the UV, [42,43] outside of our observation range.

In the first class of such scenarios, surface modification of $TiO_2$ nanoparticles by glycerol results either in the formation of a new trapping site for the electron or

21.

elimination of a subset of such trapping sites. In particular, EPR studies [3] suggest that surface modification of TiO$_2$ nanoparticles results in the elimination and/or modification of electron traps at the nanoparticle surface. Thus, a possible way to interpret our results is to assume that the TA spectrum observed in glycerol solutions (trace (i) in Fig. 2) is from the interior electrons whereas the spectrum in aqueous solutions without hole scavenger (trace (iii) in Fig. 2) is a composite spectrum originating from both the interior and surface-trapped electrons.

First, we consider the possibility that the TA spectrum in the aqueous solution (without glycerol) is composite. If this were the case, two unlikely coincidences should occur, namely (i) the decay kinetics for different electron species should be identical over many decades in time (as otherwise the TA spectrum would be time dependent), and (ii) the photobleaching of these species by 532 and 1064 nm light should be exactly the same (as otherwise selective photobleaching of one of these species would occur). The only way in which such a situation might occur naturally would involve rapid equilibrium involving these different electron species (that would have to occur well within the duration of the photoexcitation pulse). As mentioned in the Introduction, for some electron species (e.g., the CB electron and electrons in shallow traps) such a rapid equilibration is quite likely. [16,18] To account for our data (assuming that the TA spectrum is from the electrons only) it is necessary to postulate that the TA spectrum observed in the aqueous solution is either from a *single* electron species or several such species in rapid equilibrium. The spectral transformation occurring in the glycerol solutions has to be interpreted as evidence that the *modification* of TiO$_2$ nanoparticles by glycerol leads to the formation of a new type of electron trap or, alternatively, to the elimination of certain surface traps which shifts the equilibrium between the light-absorbing electron species contributing to the TA spectrum in the aqueous solution.

We turn, therefore, to the scenario, in which the modification of the TiO$_2$ surface by glycerol results in a *new type* of the electron trap. [3,25,26] The electrons descend into such traps in preference to the ordinary traps at the surface of a TiO$_2$ nanoparticle in the aqueous solution. To distinguish between the corresponding electron species, we will denote the electron in the regular trap as $e_w^-$ and the electron in the polyol-modified trap

22.

as $e_g^-$. The "instantaneous" transformation of the TA spectrum in Fig. 4b would be interpreted as evidence for the competition between these two traps for the CB electron. The spectral transformation observed in the first 200 ns would be indicative of the slow $e_w^- \rightarrow e_g^-$ transfer from the shallower to the deeper trap. In such a case, each intermediate spectrum in Fig. 4b would be a weighted sum of the initial spectrum (trace (i)) from $e_w^-$ and the final spectrum (trace (ii)) from $e_g^-$. This is indeed the case (section 3.1). While this interpretation would be consistent with our observations, it poses three problems. First, despite their different nature, $e_w^-$ and $e_g^-$ would be required to have *exactly* the same spectra for $\lambda > 900$ nm. Second, it seems unlikely that $e_g^-$ (which is the electron in a *deeper* trap) has an absorption band to the red of $e_w^-$ (which is the electron in a *shallower* trap). Third, absorption spectra similar to traces (ii) and (iv) in Fig. 2 have been observed from reduced $TiO_2$ nanoparticles and films containing no glycerol. (e.g., refs. 33 and 34). Thus, this scenario is unsatisfactory. [44]

We are left, therefore, with a scenario in which binding to glycerol eliminates a subset of surface traps and thereby shifts the equilibrium between the light-absorbing electron species, one of which has the absorption spectrum (mis)identified in section 3.1 as that from the trapped hole. There are data in the literature that seemingly support such a scenario: For example, it has been observed that electrochemical reduction of thin anatase films in a certain range of potentials yields an electron species (see Fig. 2 in ref. 22) whose spectrum resembles the TA spectrum shown by trace (ii) in Fig. 2. This scenario, however, also poses problems, because not only the prompt but also the "final" spectrum (attained at 200 ns) would depend on the glycerol concentration since the equilibrium between the electron species would depend on the availability of unmodified surface traps which, in turn, would depend on the extent of binding of these traps to glycerol. A related problem is how to explain the slow transformation of TA spectra in Fig. 4b and the results of Fig. 11. Since the postulated equilibrium between different electron species has to be rapid, the slow process can only involve transformations of the traps themselves while the electron resides in these traps. While it is possible to envision such transformations, their thermodynamics and kinetics would strongly depend on the

23.

structure of the surface modifier. Since exactly the same "final" spectrum is observed for all polyols and carbohydrates, [31] including PVA [32], this scenario is also unsatisfactory. Lastly, it is difficult to account for the fact the spectra very similar to these "final" spectra have been observed from reduced $TiO_2$ nanoparticles and films containing no hole scavengers or surface modifiers. [33,34]

As seen from our examination, all scenarios involving elimination or modification of electron traps appear to run into one problem or another. There exists, however, a class of the two-electron scenarios that avoids these problems. Specifically, it may be postulated that the presence of the hole on the same nanoparticle modifies the absorption spectrum of the electron in the visible. In other words, the spectrum given by trace (iii) in Fig. 2 is from a self-trapped exciton, whereas trace (i) originates from the decoupled electron which resides on a $TiO_2$ nanoparticle without the hole. The dispersive recombination of electron-hole pairs (Fig. 1) is then viewed as the collapse of the exciton. As shown in Fig. 7S, when the TA spectrum in water is plotted as a function of photon energy $E_{ph}$, the spectrum for $E_{ph} > 1.25$ eV is Gaussian, as would be expected for an exciton-like species. It is easy to see that all statements made in sections 3 and 4.1 can be restated in the exciton model by renaming the "light-absorbing hole" or "electron-hole" pair as "trapped exciton", because any process that depletes the hole automatically converts one kind of spectrum to another. In other words, our results appear to be equally compatible with the existence of a light-absorbing hole or a self-trapped exciton in aqueous $TiO_2$ nanoparticles. Below, we consider both of these possibilities in more detail.

*4.3. Light-absorbing holes: pro and contra.*

A photogenerated hole absorbing light in the visible would account for our observations (sections 3 and 4.1). Such a rationale is not without a precedent, as there have been previous reports of such holes from photolyzed $TiO_2$ nanoparticles and films, both on the picosecond [36] and nanosecond time scales. [10,11,12] Perhaps, the strongest evidence of holes absorbing in the visible has been given by Bahnemann et al. [10,11,12] who observed a TA spectrum from photoexcited aqueous $TiO_2$ nanoparticles impregnated with small islands of metallic platinum ("platinized" nanoparticles). The *Pt*

24.

patches rapidly remove photogenerated electrons from the TiO$_2$ nanoparticles, leaving behind the hole. The resulting species oxidizes $Br^-$ to $Br_2^-$ and has an absorption spectrum with an onset at 700-800 nm [12] and a maximum around 400 nm [10] or 470 nm. [12] This spectrum (Fig. 5b) bears no resemblance to the spectra of "holes" generated in pulse radiolysis (see below). On the other hand, this spectrum is similar to the one shown in Fig. 1, trace (ii) and Fig, 5b, trace (i). Despite these observations, the general consensus is in favor of a *UV-absorbing* hole. Two arguments supporting this may be given:

First, the reaction of TiO$_2$ nanoparticles with strong oxidizing radicals, such as radiolytically-generated $SO_4^{\bullet-}$ and $HO^{\bullet}$ radicals, yields a species that absorbs in the UV, with an onset below 400 nm. [42,43] Assuming that these radicals react with the nanoparticles by electron transfer, it may be inferred that the holes do not absorb in the visible. To account for the TA spectrum of the tentative holes observed by Bahnemann et al. [10,11], Lawless et al. [42] suggested that this spectrum originated from the photoexcitation of *Pt* islands rather than TiO$_2$ nanoparticles themselves. This argument does not seem to be supported by the results of the present study, in which the absorbance in the visible similar to that reported by Bahnemann et al. [10,11,12] was observed in the absence of platinum. We may further object that the pulse radiolysis studies do not provide decisive evidence that electron transfer between these radicals and the TiO$_2$ nanoparticle indeed occurs (as also noted by Lawless et al.) [42] The occurrence of the electron transfer, with a nearly diffusion controlled rate, [42,43] seems unlikely as the standard potentials for the sulfate and hydroxyl radicals are ca. 0.9 eV and 0.45 V, respectively, lower than this potential for holes in hydrated anatase. [43]

It is possible that the reaction of hydroxyl radical with a TiO$_2$ nanoparticle yields $HO_{ads}^{\bullet}$ as the main product (the same applies to the $SO_4^{\bullet-}$ radical). As mentioned in the Introduction, it is presently believed that at least two hole species, the mobile hole (observed in our TA experiments) and $HO_{ads}^{\bullet}$, are generated by UV photolysis of aqueous TiO$_2$ nanoparticles. [27,28,29] The latter species is a minor product with a quantum yield ca. 100 times lower than that of the more abundant mobile hole. [27] Pulse radiolysis data alone are insufficient to determine which reaction, surface addition or electron transfer, occurs when these radicals encounter a nanoparticle in the water bulk. A convincing demonstration for the occurrence of electron transfer would be the similarity of reaction patterns and rates for radiolytically and photolytically generated holes. Such a demonstration is presently lacking, and this absence of supportive results weakens the whole argument favoring the UV-absorbing holes.

Second, the $O\ 2p$ radicals of the $Ti^{IV} - O^{\bullet}$ type identified by EPR [3,26] as possible hole centers are not expected to be strong absorbers in the visible. The obvious objection to this argument is that the light-absorbing hole might be a different type of oxygen hole center. In particular, if the hole were trapped on the bridging oxygen of a $Ti^{IV} - O - M^{n+}$ center, where $M^{n+}$ is the (impurity) metal cation, the crystal field of this cation would split the degenerate $2p$ orbitals, resulting in the absorption of visible light by the hole center. [45] A classical example of such an impurity center is the aluminum hole center in smoky quartz ($SiO_2$:Al). In this solid, overcoordinated $Al^{III}$ ion substitutes for $Si^{IV}$, providing a negatively charged trap for the hole. While the $Si^{IV} - O^{\bullet}$ center does not absorb visible light, $Si^{IV} - O^{\bullet}...Al^{III}$ centers strongly absorb across the visible. [45] Note that an impurity content of just 7 ppm would be sufficient to have one metal ion per $TiO_2$ nanoparticle in our photosystem (if all such ions are in the $TiO_2$ phase). The typical concentrations of impurity ions in our colloidal solutions were much higher (> 40 ppm). Thus, a commonly occurring bi- or tri- valent impurity ion, such as $Al^{III}$, might be the sought-after progenitor of the oxygen hole center. Since such an ion would substitute for tetravalent $Ti^{IV}$, the precursor of the hole would be a negatively charged trap. Such a trap can be a deeper trap than $Ti^{IV} - O^-$ or $Ti^{IV} - OH$ groups at the surface that are believed to be the precursors of the $Ti^{IV} - O^{\bullet}$ hole. If this impurity ion is diamagnetic and has small nuclear dipole moment, no additional fine or hyperfine structure would be observed in the EPR spectra of the resulting hole center. Given the high crystallinity of anatase nanoparticles, [2,38] it is likely that such an impurity ion would be at or near the nanoparticle surface. Since the partition of (impurity) ions between the aqueous and $TiO_2$ phases would be *pH* dependent, sensitivity of the TA spectrum to *pH* of the solution [32] finds a natural explanation.

While our identification of photogenerated holes with impurity-related oxygen hole centers is tentative, this suggestion may be verified without going to the extreme effort of eliminating the ubiquitous impurity. The unpaired electron on the oxygen would be coupled to the nucleus of the impurity ion by weak dipole interaction. Such an interaction may be observed using electron spin echo envelope modulation (ESEEM) and electron nuclear double resonance (ENDOR) spectroscopies.

*4.4. Self-trapped exciton.*

As discussed in section 4.2, the alternative explanation of our results would be admitting that the holes in the aqueous $TiO_2$ nanoparticles do not absorb in the visible, and the vis spectrum shown in Fig. 2, trace (iii) is from a self-trapped exciton. We are aware of no evidence for the existence of such excitons in aqueous $TiO_2$ nanoparticles, although there is ample evidence that self-trapped excitons are formed in crystalline anatase [46] and anatase films. [47] In these materials, self-trapped excitons decay by radiative recombination and yield photoluminescence spectra centered at 2.4 eV.

The chief problem is to explain the *absorption* spectrum of the tentative exciton. The exciton in crystalline and thin-film anatase contributes to the smooth Urbach tail at the band edge; [46] these excitons do not yield the absorption spectrum shown in Fig. 2, trace (iii). Furthermore, one would expect that strong electron-hole coupling would be readily identifiable in the EPR spectra of UV illuminated $TiO_2$ nanoparticles. No evidence for such a coupling has been found. Thus, the self-trapped exciton scenario, while plausible, lacks experimental support.

*4.5. Photostimulated electron detrapping.*

The very fact that photoexcitation of electrons results in their rapid recombination with trapped holes suggests their identity as *trapped* charges. A claim is sometimes found in the literature that the 700-900 nm band of the trapped electron is actually from free CB electrons. It is difficult to explain how such a free carrier could have existed on the same $TiO_2$ nanoparticle with the hole over many microseconds, to give the composite spectrum shown in Fig. 2, trace (iii). As pointed out in the Introduction, the view that the 700-900 nm absorbance originates from free electrons appears to be inconsistent with many observations. Ultrafast kinetic measurements [4] suggest a subpicosecond time scale for free electron trapping. THz spectroscopy of dye-sensitized $TiO_2$ films [8] suggests negligible extension of the CB electron spectrum to the IR and points to its polaronic nature. Microwave conductivity studies [17,18,20] suggest that the electrons in $TiO_2$ *nanoparticles* reside in shallow traps near the CB edge. Such shallow-trap electrons would readily account for the diffuse IR absorbances observed in refs. 14 and 16. Indeed, with a few exceptions [48] all observations of persistent "free electrons" in titania nanoparticles can be interpreted as evidence for the existence of such shallow-trap electrons. As considerable delocalization of the electron wavefunction is likely in these shallow-trap states, evidence for such a delocalization (e.g., the recent observation of Stark effect on the Ti-OH vibrations) [49] does not necessarily indicate the involvement of the *free* electron.

In this regard, it is noteworthy that the quantum yield for electron photobleaching decreases by an order of magnitude from 532 nm (2.33 eV) photoexcitation to 1064 nm (1.17 eV) photoexcitation (section 3.2). Only electrons in deep traps (0.5-1 eV) would account for such a precipitous decrease in the photobleaching efficiency. The highly asymmetrical shape of the electron spectrum (trace (i) in Fig. 7S) also suggests that this spectrum is composite. Specifically, it appears that a broad absorption line centered at ca. 1.5 eV is juxtaposed on a diffuse absorbance that increases as $\lambda^{1.5-2}$ towards the red (this absorbance may be the blue extension of the IR spectra observed by Hoffmann et al. from oxygen-treated photoilluminated rutile $TiO_2$ surfaces). [16] These two spectral contributions might correspond to the electrons in deep and shallow traps, respectively. As discussed in section 4.2, a composite nature of the electron absorbance would be compatible with the observations on the nanosecond time scale provided that the equilibration of different trapped states occurs on a much shorter time scale. Unfortunately, the existing picosecond data for unmodified $TiO_2$ nanoparticles are limited to a narrow spectral interval between 400 and 700 nm where both the electron and hole absorb probe light (section 3.1). While there is considerable spectral evolution within the first 50-100 ps, [4,36,37] this evolution can be interpreted in many different ways.

Our spectral data hint that at some point, as the wavelength of the analyzing light increases towards the IR, there should be a gradual transition from a TA spectrum dominated by electrons residing in deep traps (which add little to the dc and ac conductivity) to a TA spectrum dominated by electrons in shallow traps (responsible for the electron conduction). Perhaps, such a transition occurs somewhere in the mid IR, where the electron absorption in aqueous $TiO_2$ nanoparticles cannot be observed due to the strong extinction of the IR light by the solvent.

It is worth mentioning that there are results in the literature which appear to contradict the notion that different trapped-electron species on the same $TiO_2$ nanoparticle rapidly equilibrate. Recently, Dimitrijevic et al. [38] used pulse radiolysis to study the reaction of $TiO_2$ nanoparticles with hydrated electrons. The resulting TA spectra did not look like the spectra from "persistent" electrons obtained by Safrany et al. [34] using the same method for similar nanoparticles. The electron absorbance for $\lambda > 1.2$ µm was flat; [38] there was also a bell-shaped absorption band with a maximum at 680 nm. Full coverage of the nanoparticle by dopamine selectively removed that bell-shaped band, leaving a spectrum which was flat over the entire visible and near IR. One would expect that the absorption spectrum of the "electron" does not depend on the method used for its generation, be it photolysis, pulse radiolysis, or electrochemical reduction. While it is possible that different electron species are generated in photolysis and radiolysis, the postulated rapid equilibration would eliminate any such differences. As explained in sections 3.2 and 4.2, coexistence of several distinctive light-absorbing electron species on the same nanoparticle, in the absence of the rapid equilibration, is not supported by our results.

**5. Conclusion.**

It is shown that 532 nm and 1064 nm laser photoexcitation of trapped electrons generated in 355 nm photolysis of aqueous titania ($TiO_2$) nanoparticles (diameter 4.6±0.5 nm, *pH*=4) causes rapid photobleaching of their absorbance band in the visible and near IR. This photobleaching occurs within the duration of the laser pulse (6 ns FWHM); it is caused by photoinduced electron detrapping followed by recombination of the resulting free electron and a trapped hole. The quantum yield for the electron photobleaching is ca. 0.28 for 532 nm and ca. 0.024 for 1064 nm photoexcitation. The drastic reduction in the quantum efficiency with increasing wavelength of the photoexcitation light suggests that the absorption spectrum of the electron in the near IR originates both from the electrons in shallow traps and the electrons in deep traps; there should be rapid (< 1 ns) equilibration between all such states.

Complete separation of the spectral contributions from trapped electrons and holes is demonstrated using glycerol as a selective hole scavenger (Fig. 2). The holes absorb across the entire visible; their absorption increases towards the UV. The electron absorption increases towards 900-1000 nm and then slowly decreases towards the IR. The characteristic 650 nm band of photoilluminated aqueous $TiO_2$ nanoparticles is composite. Our results (as well as the results of Bahnemann et al. [10,11,12]) indirectly suggest that holes obtained by reaction of oxidizing radicals (such as hydroxyl) with $TiO_2$ nanoparticles [42,43] have absorption properties different from those of the photogenerated holes. We speculate that the vis-light absorbing *O 2p* holes originate from a common ion impurity in colloidal $TiO_2$. This suggestion can be verified using advanced magnetic resonance spectroscopies. There is an intriguing possibility that $TiO_2$ nanoparticles can be doped by this ion on purpose to facilitate hole trapping and improve photocatalytic efficiency.

In the absence of the hole scavenger (glycerol), no evolution of the transient absorbance spectra (400 to 1350 nm) was observed out to 10 μs, both in $N_2$- and $O_2$-saturated aqueous solutions. In the presence of 5 vol % glycerol, ca. 45-50% of the light-absorbing holes are scavenged promptly within the duration of the 355 nm photoexcitation pulse, another 40-45% of the holes are scavenged at a slower rate over the first 200 ns after the photoexcitation pulse, and the remaining 5-10% are not scavenged, even at the high concentration of the scavenger (>10 vol %). A reaction with chemi- and physi- sorbed glycerol can account for the prompt and the slow hole decay, respectively. The mechanism for this reaction is further discussed in the companion paper. [31]

Our results can also be explained by a mechanism in which the absorption spectra of electrons with and without the hole present on the same $TiO_2$ nanoparticle are different (sections 4.2 and 4.4). This interpretation can be construed as evidence for the formation of self-trapped excitons in the anatase nanoparticles. However, at the present time this explanation has little experimental support. Other scenarios in which the composite spectrum shown in Fig. 2 originates from two or more electron species (rather than the electron and the hole) also seem to be excluded by our results (section 4.2).

## 6. Acknowledgement.

I. A. S. thanks Drs. T. Rajh, P. Kamat, and D. Meisel for stimulating discussions and Drs. N. Dimitrijevic and Z. V. Saponjic for the preparation of some $TiO_2$ samples. This work was performed under the auspices of the Office of Science, Division of Chemical Science, US-DOE under contract number W-31-109-ENG-38.

**Supporting Information Available:** A single PDF file containing Figures 1S to 7S with captions. This material is available free of charge via the Internet at http://pubs.acs.org.

**Figure captions.**

**Fig. 1**

*Solid lines:* Typical decay kinetics of transient absorbance (TA) detected at $\lambda = 700$ nm following 355 nm photoexcitation of $TiO_2$ nanoparticles in $N_2$-saturated water (3 ns FWHM laser pulse, incident photon fluence of 64 mJ/cm$^2$; 1.35 mm optical path cell, 24% transmission of 355 nm light). This kinetic trace is the average of 40 laser shots; the sample was flowed at 1.5 cm$^3$/min during the acquisition at 1 Hz. ). See section 2 for the details of the setup and experimental procedures. The TA signal at 700 nm originates from the absorbances of trapped electrons and holes (ca. 1:1); the decay is due to recombination of the electrons and holes residing on the same $TiO_2$ nanoparticle. Under these photoexcitation conditions ca. 1-2 pairs are generated per nanoparticle. In (a), the first 300 ns of these kinetics are shown on a linear scale, in (b), the kinetics is shown on the double logarithmic scale (out to 8 μs). The dashed lines in the same plots correspond to a power law $t^{-\alpha}$ dependence with $\alpha \approx 0.46$. Using our setup, this power-law decay can be traced out to 50-100 μs. Time profiles of TA kinetics for other wavelengths of the analyzing light are shown in Fig. 1S (450-950 nm) and Fig. 2S (950-1350 nm) in the Supplement. All of these kinetic profiles are identical within experimental error, and the shape of the TA spectrum is independent of the delay time (Fig. 2).

**Fig. 2**

*Open symbols:* Normalized TA spectra from aqueous $TiO_2$ nanoparticles photoexcited by 355 nm laser light. These spectra were obtained using a fast Si photodiode. The integration windows are, respectively, 11-13 ns *(open circles),* 13-20 ns *(open squares),* 20-50 ns (open upright triangles), and 50-170 ns *(open triangles).* As seen from this plot, the spectrum does not change with the delay time (this can also be seen from the data of Figs. 1S and 4a). *Filled diamonds:* near-IR spectrum from the same photosystem obtained using a fast Ge photodiode (50-200 ns integration). The TA spectra in the near IR also do not depend on the delay time (see Fig. 2S). Traces (i) to (iii) demonstrate the decomposition of the TA spectrum in the visible into the contributions from trapped electrons and holes. Trace (i) is the normalized TA spectrum at 200-370 ns from $TiO_2$ nanoparticle solution containing 5 vol % hole scavenger (glycerol). This TA spectrum is mainly from the electrons (some absorbance is from holes in the nanoparticle interior that cannot be scavenged by glycerol). Trace (ii) is the normalized spectrum of the light-absorbing holes (see Fig. 5 b). The sum of these two absorptions gives trace (iii) *(crosses)*. Trace (iv) gives the electron absorbance only (see Fig. 8b). The solid lines are guides for the eye.

**Fig. 3**

Time evolution of TA spectra (400-1000 nm) obtained under identical photoexcitation conditions in oxygen saturated (a) aqueous solution of $TiO_2$ nanoparticles and (b) the same solution containing 5 vol % of glycerol. The integration windows are, from top to bottom, 9-13 ns *(open circles),* 13-23 ns *(filled squares),* 23-45 ns *(open squares),* 45-75 ns *(filled triangles),* 75-120 ns *(open triangles),* 120-200 ns *(filled diamonds),* and 200-370 ns *(open diamonds),* respectively. For $t > 300$ ns, the spectral evolution is negligible. The time evolution of the near IR spectrum (800 to 1350 nm) for glycerol solution is shown in Fig. 3S.

**Fig. 4**

TA spectra from Fig. 3 normalized by the absorbance signal at 900 nm (same symbols as in Fig. 3; see the legend in (a)). In (b), trace (i) is the prompt TA spectrum from (a) and trace (ii) is the "final" TA spectrum attained at 200-370 ns. The solid lines are guides to the eye. While in the absence of the hole scavenger there is no spectral evolution with the delay time, addition of the hole scavenger results in the removal of the absorbance in the visible (which originates from the hole). As seen from the plot in (b), ca. 45% of the hole absorbance is removed during the laser pulse; another 45-50% is removed over the first 200 ns. The residual absorbance at $t > 200$ ns is mainly from trapped electrons. See also Fig. 4S, which gives the TA kinetics from the 5 vol % glycerol solution that are normalized at 200 ns. The short-lived TA signal in the visible is from the holes.

**Fig. 5**

(a) *Symbols:* difference TA spectra (same coding as in Fig. 4) obtained by subtraction of the TA spectrum at 200-370 ns from other traces (excepting curve(i)) in Fig. 4b (5 vol % glycerol solution). The spectrum at 200-370 ns *(open diamonds)* is shown by trace (ii); trace (i) *(open circles)* is the same as trace (i) in Fig. 4b, i.e., it is the TA spectrum from the aqueous solution that contains no hole scavenger. The solid lines are guides to the eye. *Filled circles* give a weighted sum of trace (ii) and the prompt TA spectrum of the hole *(open circles)*. As seen from the comparison with trace (i), the TA spectrum observed in the aqueous solution can be accurately represented as the sum of these two spectral components (see also Fig. 2) (b) Same as (a); the difference spectra were normalized to illustrate that the spectrum of the species that is "removed" by glycerol over the first 200 ns after 355 nm photoexcitation does not change with the delay time (trace (i)). For comparison, trace (ii) from (a) is reproduced. Filled symbols are normalized TA spectra of (tentative) vis-absorbing holes on "platinized" $TiO_2$ nanoparticles studied by Bahnemann et al.: *filled circles* are from ref. 10 (Fig. 8 therein), *filled squares* are from ref. 12 (Fig. 5 therein).

**Fig. 6**

Decay kinetics of TA observed, using (a) 900 nm, (b) 600 nm, and (c) 470 nm analyzing light, from oxygenated aqueous $TiO_2$ nanoparticle solution containing 0, 1, 2.5, 5, and 10 vol % glycerol. These kinetics were obtained under identical photoexcitation conditions using the same stock solution. Only the first 370 ns after the photoexcitation are shown. As the concentration of glycerol increases so does the TA signal attained at 200-350 ns. Observe the decrease in the prompt TA signal in the visible with increasing glycerol concentration. The TA kinetics at 900 nm (where only trapped electrons absorb) gives the decay kinetics of the electron.

**Fig. 7**

The data of Fig. 6, replotted. (a) TA signals attained at 330-350 ns (at which time all the holes that can be scavenged by glycerol have been scavenged) vs. the molar concentration of the glycerol. These TA signals are for $\lambda = 470$ nm *(open triangles)*, 600 nm (open squares), and 900 nm *(open circles)*. The solid lines are least squares fits using the Stern-Volmer eq. (2); the constant $K_b$ is the same for all three dependencies. (b) The plot of functions $f_{900}(t)$ given by eq. (3) obtained from the data of Fig. 6a for glycerol concentrations given in the legend. As seen from this plot, all such functions are the same. The persistence of $f_\lambda(t)$ with [glycerol] is observed for all wavelengths $\lambda \geq 900$ nm (e.g., Fig. 5S), i.e., across the entire region where only trapped electron absorbs analyzing light. This persistence of $f_\lambda(t)$ means that the kinetics shown in Fig. 6a can always, at any concentration of the hole scavenger, be represented as a weighted sum of a constant and the kinetics in the aqueous solution that does not contain the scavenger (see section 3.1).

**Fig. 8**

(a) The TA signals at $\lambda = 470$ nm *(open circles)* and $\lambda = 600$ nm *(open squares)* plotted as a function of $\lambda = 900$ nm absorbance for several glycerol concentrations. The higher absorbances correspond to the higher glycerol concentration. These dependencies are linear, suggesting that addition of glycerol increases the TA signals in the visible (attained when the hole scavenging is complete) in the same proportion as it increases the electron absorbance at 900 nm. This means that these TA signals increase because the yield of persistent electrons at $t > 200$ ns increases. Note that these linear dependencies do not extrapolate to zero when the 900 nm absorbance tends to zero: As the concentration of glycerol increases and more persistent electrons are generated, the electron spectrum builds upon an absorption spectrum from some other species. (b) The "reconstruction" of this residual spectrum. Filled symbols show the TA signals at 470, 600, and 900 nm (at $t > 200$ ns) as a function of glycerol concentration. The vertical arrows point to the "residual absorbances" obtained by linear extrapolation of plots in (a). Trace (i) *(open squares)* is the TA signal attained at 330-350 ns in 10 vol % glycerol solutions (which does not change in shape as a function of [glycerol]). Trace (ii) is a scaled absorption spectrum of the hole reproduced from Fig 2, trace (ii). As seen from this plot, the arrows at 600 nm and 470 nm point to this spectrum. The same is observed at other wavelengths in the visible *(not shown);* i.e., the "residual" spectrum is from holes that cannot be scavenged by the glycerol, at any concentration of this scavenger. When this "residual" absorbance is subtracted from trace (i), the difference trace (iii) gives the absorption spectrum of the electron.

**Fig. 9**

*Thin solid lines:* Normalized decay kinetics for the vis-absorbing, "scavengeable" holes (functions $g_{600}(t)$ given by eq. (4)) at three concentrations of glycerol (1, 5, and 10 vol %, as indicated in the color scale). Trace (i) *(bold line)* is the normalized decay kinetics of the electron in aqueous solution with no glycerol, obtained at 900 nm (which must be the same as that of the hole in such a solution). Trace (ii) *(dashed line)* is trace (i) weighted by $\exp(-t/\tau_h)$ with $\tau_h \approx 130$ ns.

**Fig. 10**

(a) Photobleaching of the TA signal from $N_2$-saturated aqueous solution of $TiO_2$ nanoparticles observed at $\lambda=600$ nm. The electrons were photoexcited by a 532 nm laser (L2) pulse (6 ns FWHM, 40 mJ). The instant at which this pulse is fired, at the delay time $t_{21} = 24.5$ ns after the 355 nm pulse (3 ns FWHM, 10 mJ), is indicated by an arrow. Trace (i) is the $\Delta OD_{600}(t)$ kinetics (no 532 nm photoexcitation). Trace (ii) is the $\Delta OD_{600}^{L2}(t_{21};t)$ kinetics (following the 532 nm photoexcitation). Trace (iii) is the difference trace given by eq. (1). Trace (iv) (filled circles) is the inverted and normalized trace (iii) juxtaposed onto trace (i). See section 3.2 for more detail. (b) The fraction $q_\lambda$ of the TA signal photobleached by 532 nm light (same photoexcitation conditions as in (a)) plotted as a function of the wavelength $\lambda$ of the analyzing light. To facilitate the comparison, this fraction at 800 nm was taken to be unity. As seen from this plot, across the visible, $q_\lambda = const(\lambda)$ within the experimental error, i.e., the photobleaching spectrum is the same as the TA spectrum.

**Fig. 11**

The photobleaching efficiency $q_\lambda$ given by eq. (5) for $\lambda = 600$ nm vs. the delay time $t_{21}$ between the 532 nm and 355 nm lasers. All data points were obtained under identical excitation conditions. The bleaching efficiency in the aqueous solution of $TiO_2$ nanoparticles containing no glycerol was taken for unity (filled diamonds); this quantity does not change with the delay time $t_{21}$. The concentrations of added glycerol, in vol %, were 1 *(open triangles),* 2.5 *(open squares),* and 5 *(open circles)*. Filled circles indicate the data obtained for 5 vol % glycerol at $\lambda=900$ nm. In glycerol solutions, there is significant prompt decrease in the photobleaching efficiency (observed at $t_{21} = 25$ ns) with the increasing glycerol concentration. This decrease correlates with the fraction of holes promptly scavenged by glycerol during 355 nm photoexcitation. The decrease in the photobleaching efficiency at later delay times correlates with the "slow" hole scavenging (Figs. 5 and 9).

**References.**

\* To whom correspondence should be addressed: *Tel* 630-252-9516, *FAX* 630-2524993, *e-mail:* shkrob@anl.gov.

suggests little more than that the latter estimate is incorrect. The impedance studies of rutile *single crystals* suggest surface densities that are an order of magnitude larger; the trap density for a TiO$_2$ *nanoparticle* could only be higher.

49. Szczepankiewicz, S. H.; Moss, J. A.; Hoffmann, M. R. *J. Phys. Chem. B* **2002**, *106*, 7654.

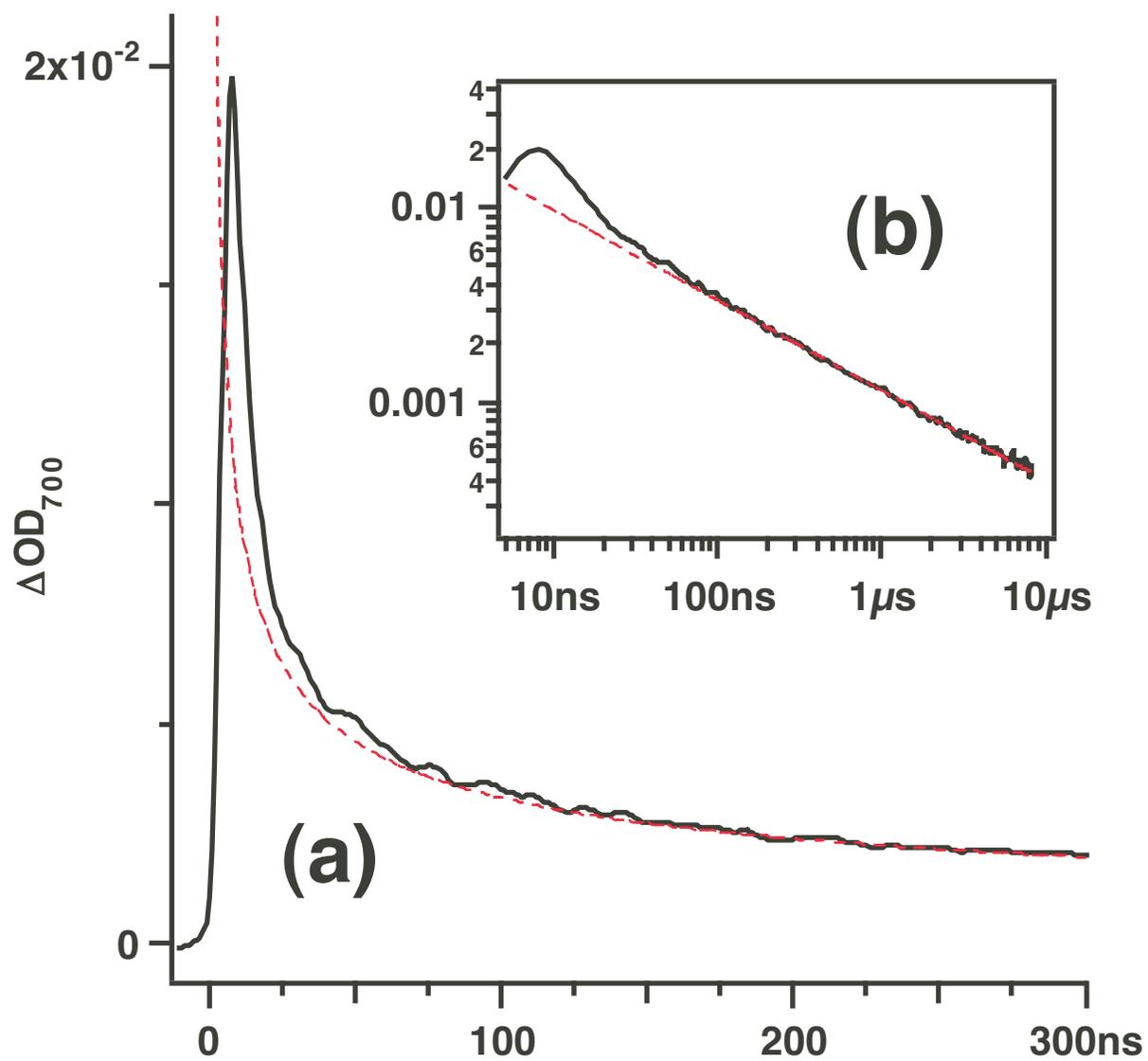

**Fig. 1; Shkrob et al.**

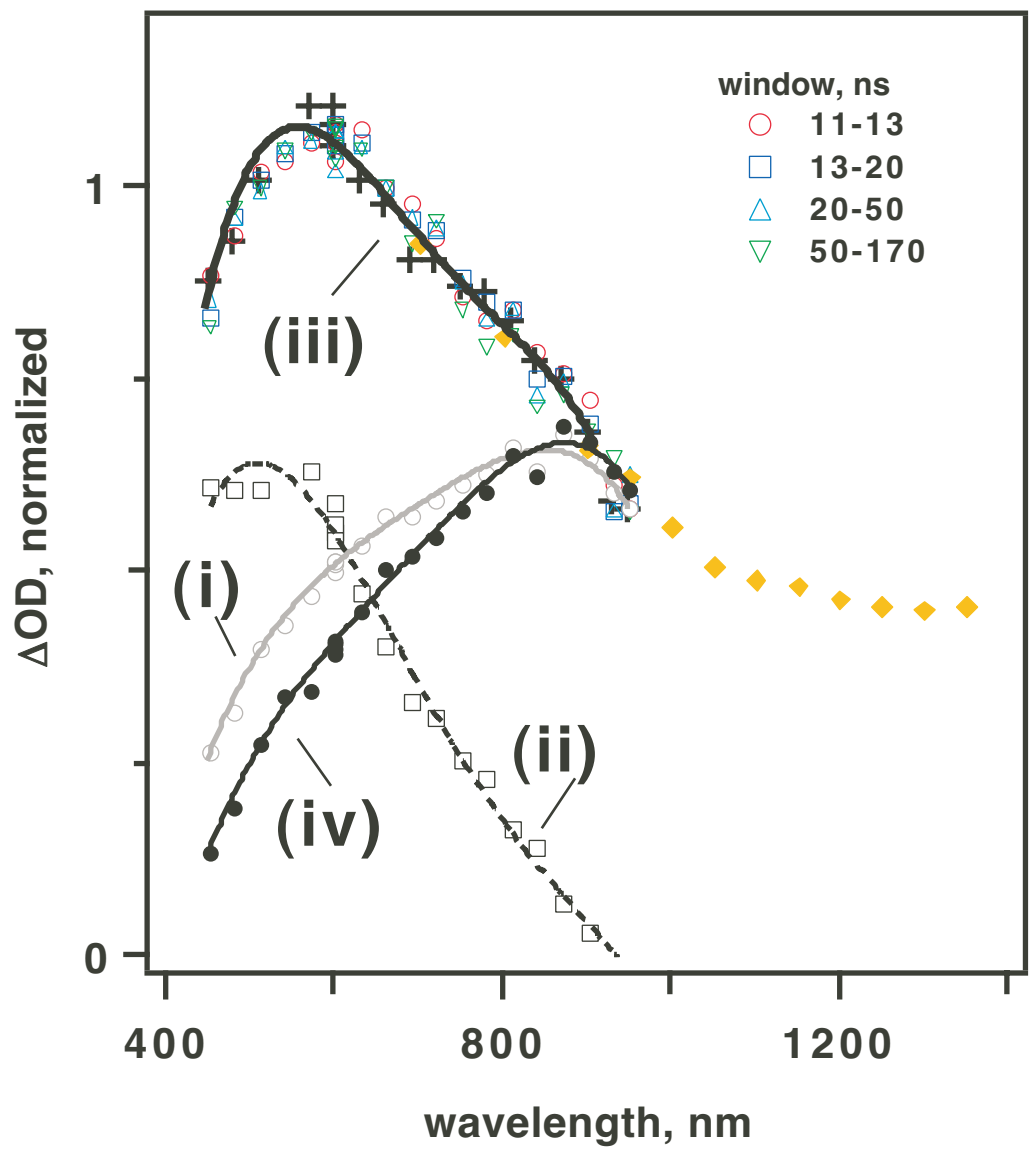

Fig. 2; Shkrob et al.

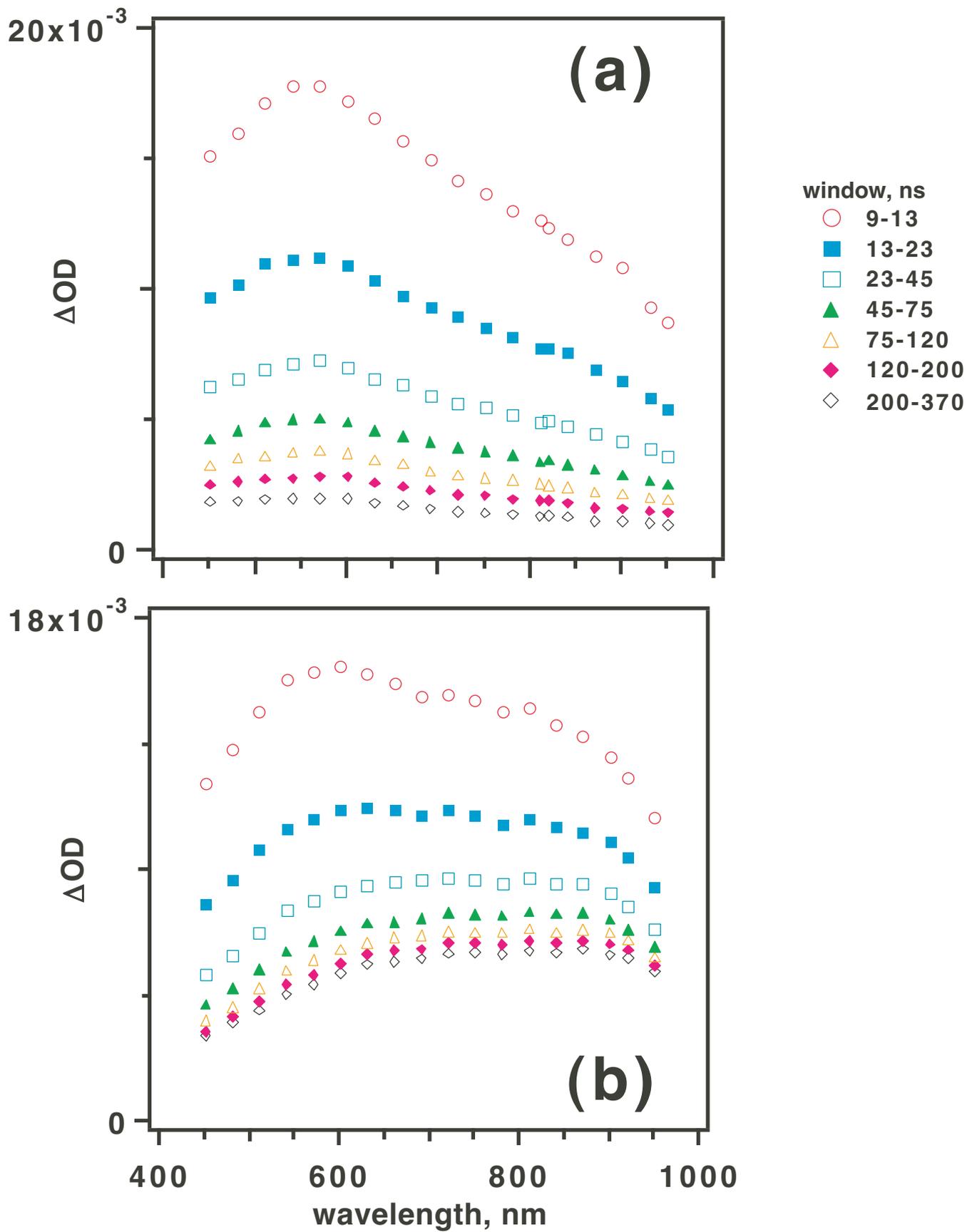

Fig. 3; Shkrob et al.

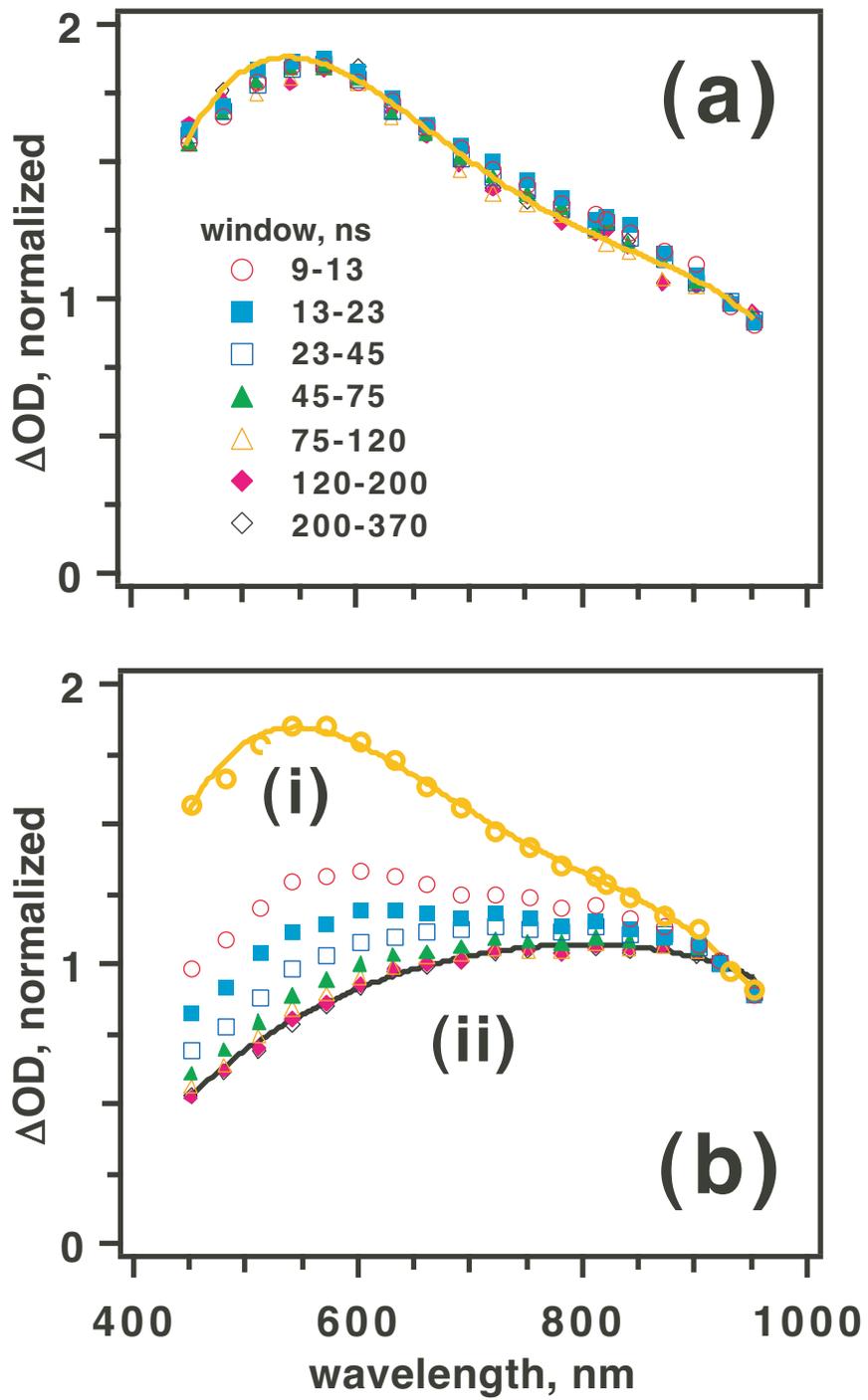

Fig. 4; Shkrob et al.

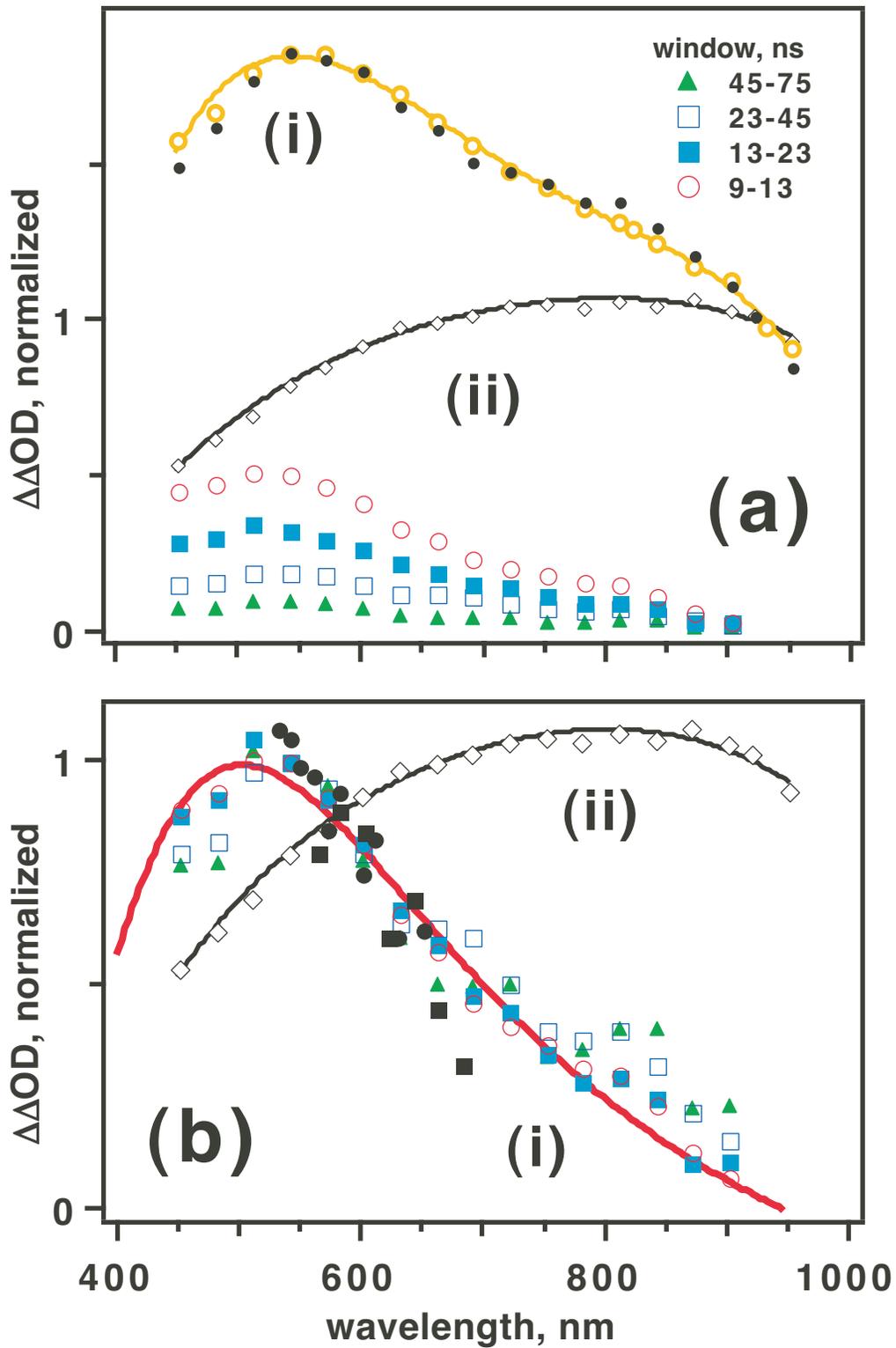

Fig. 5; Shkrob et al.

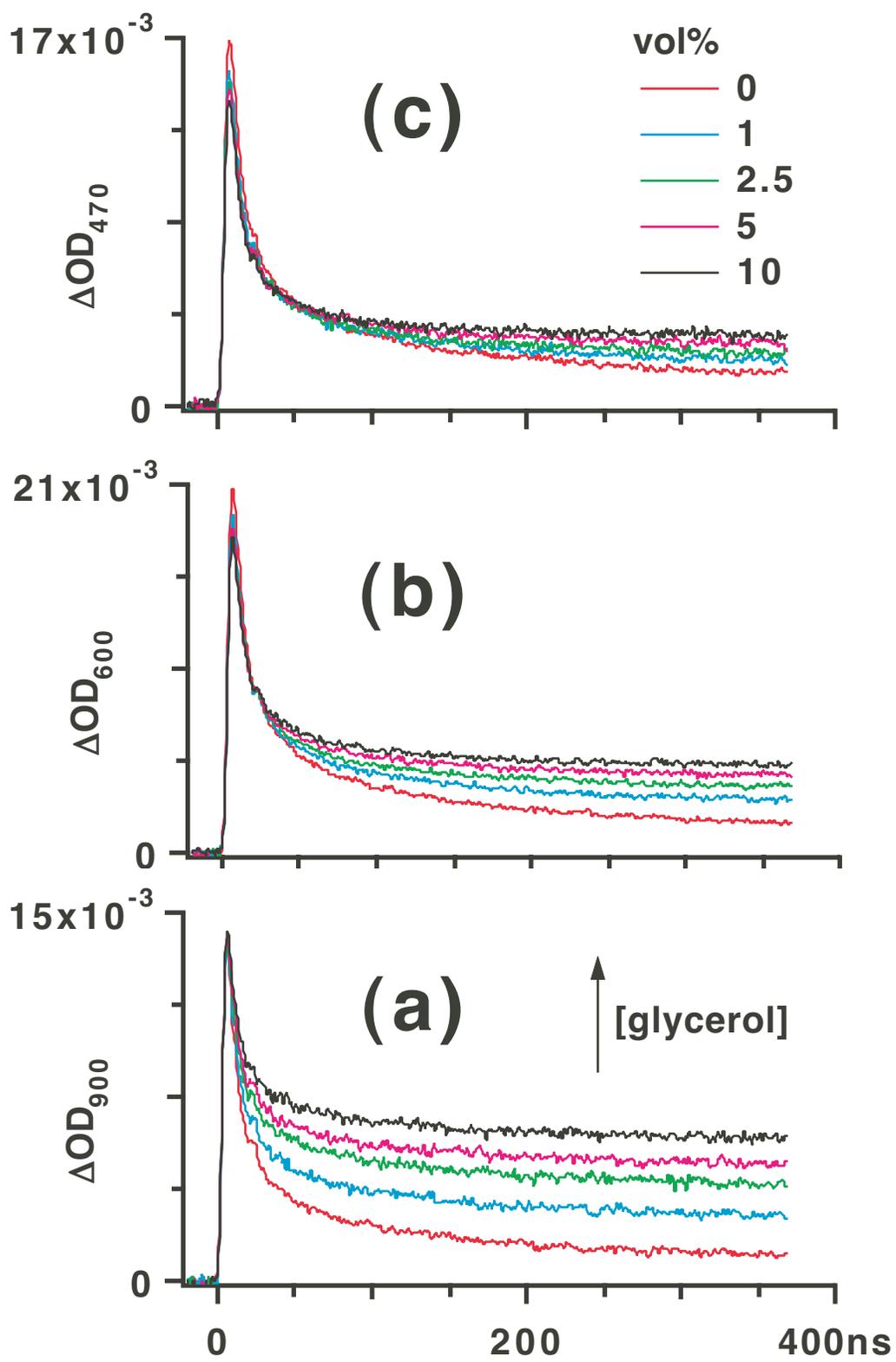

Fig. 6; Shkrob et al.

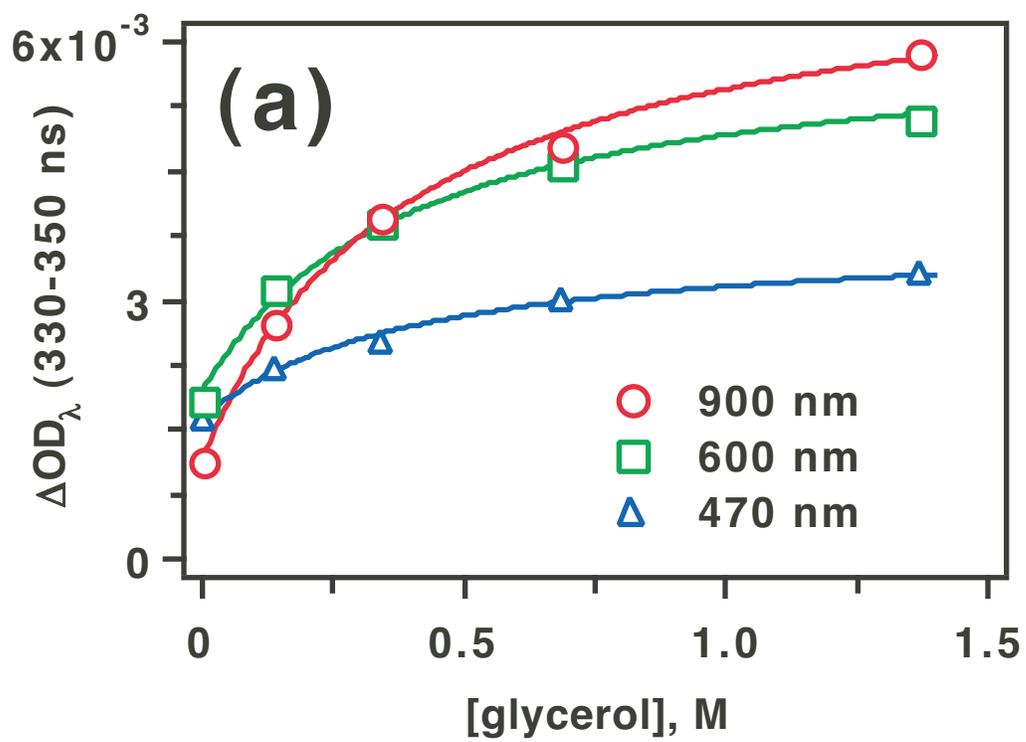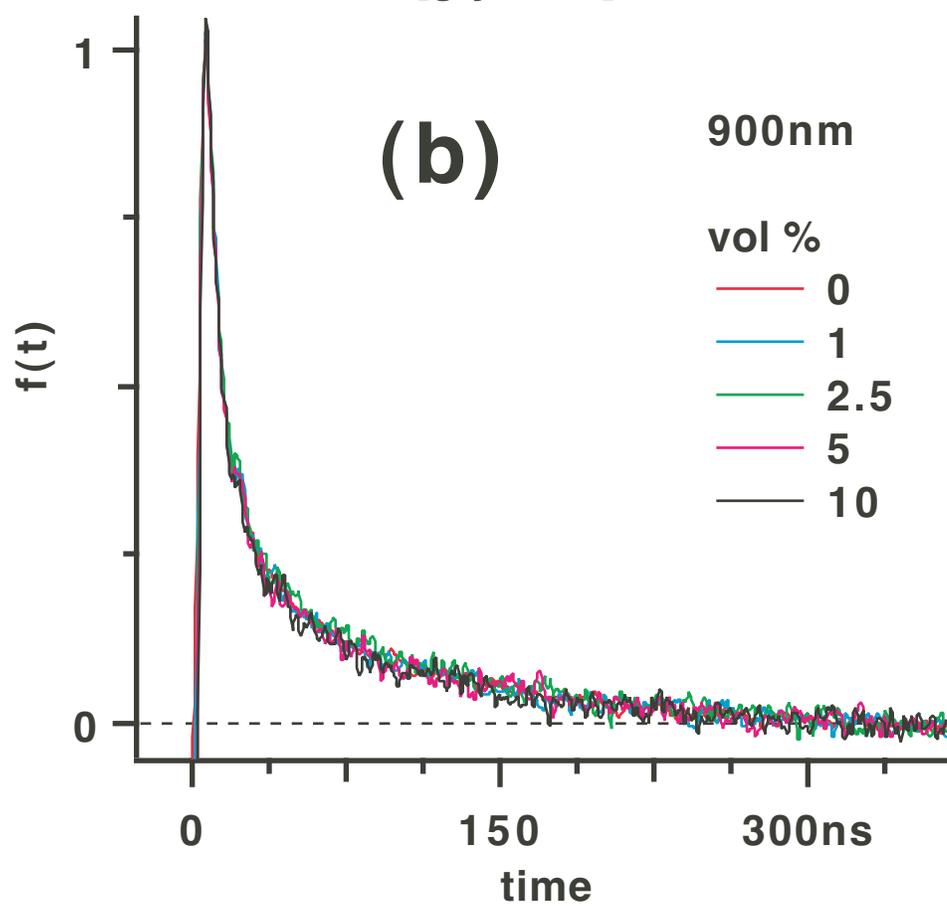

Fig. 7; Shkrob et al.

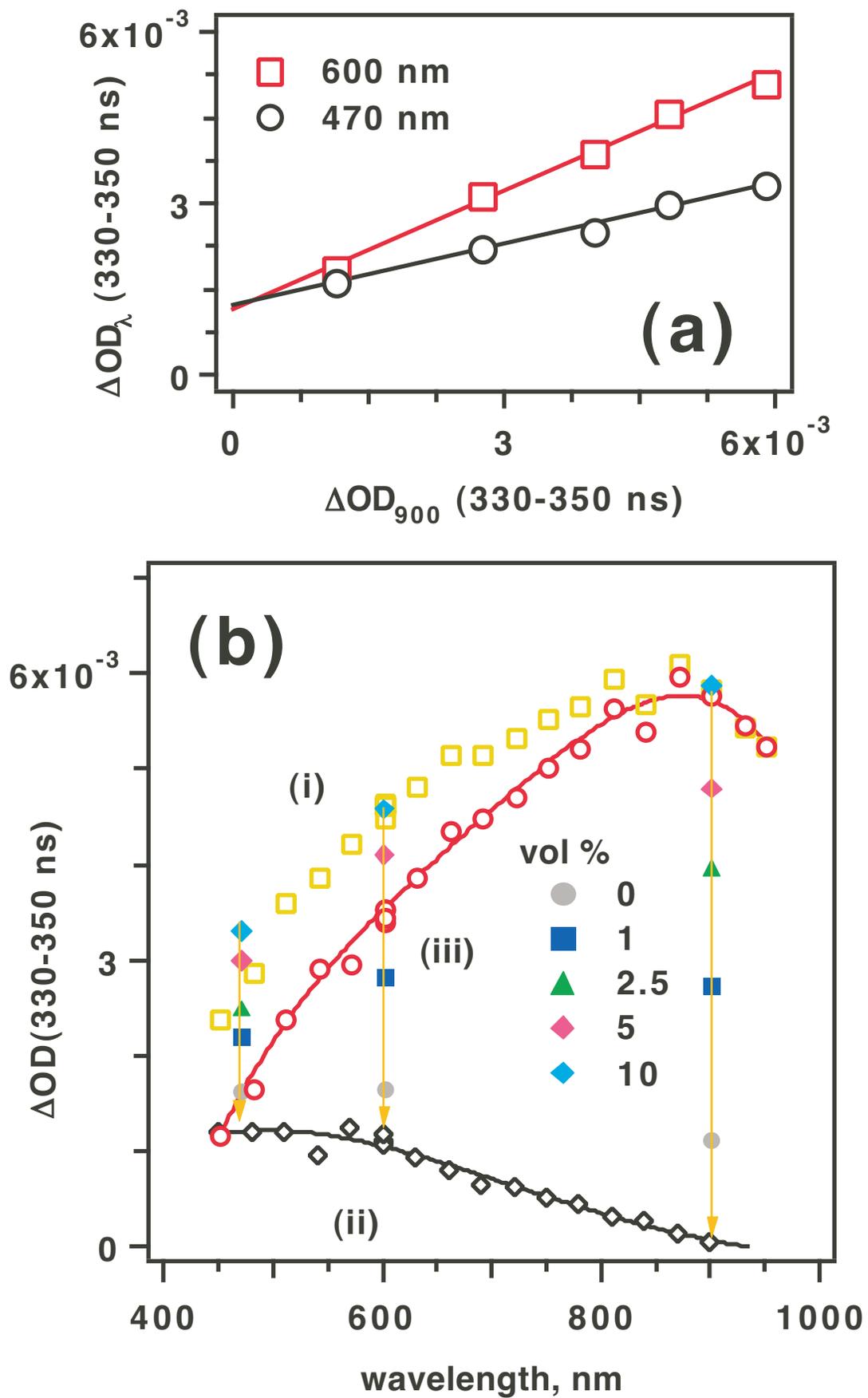

Fig. 8; Shkrob et al.

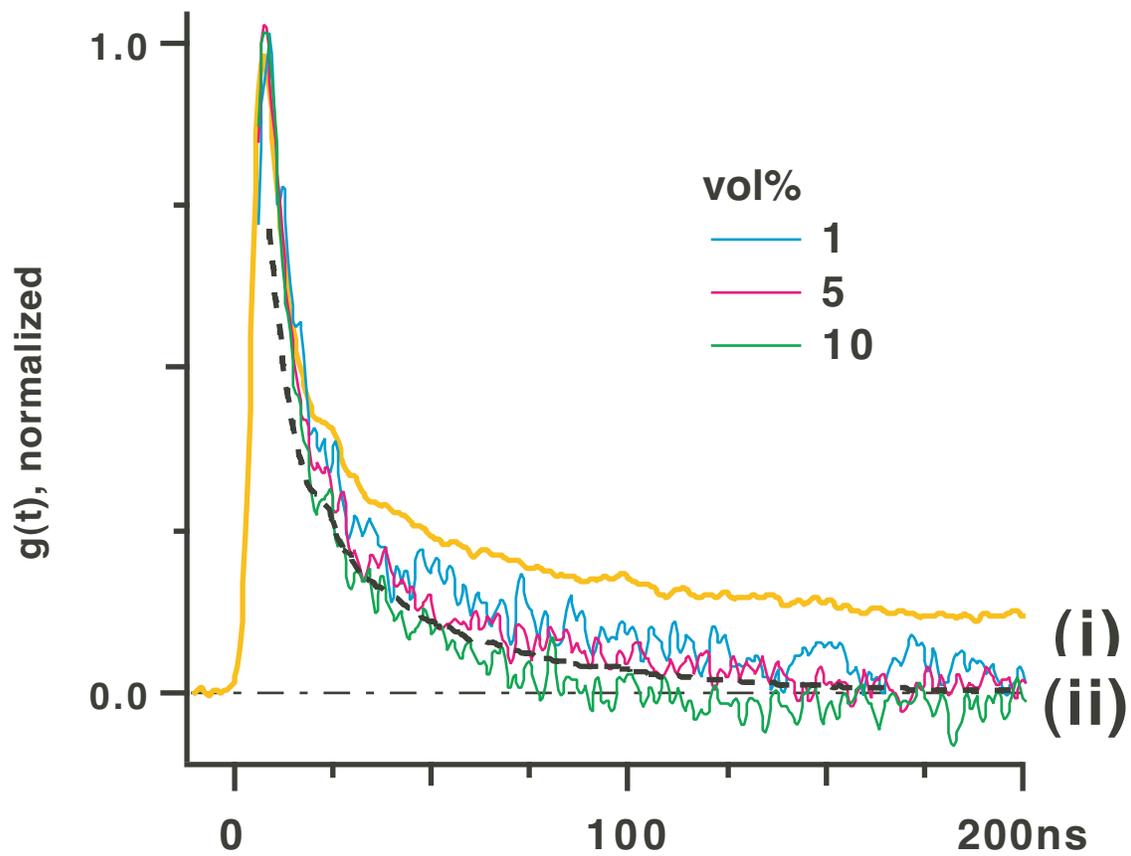

Fig. 9; Shkrob et al.

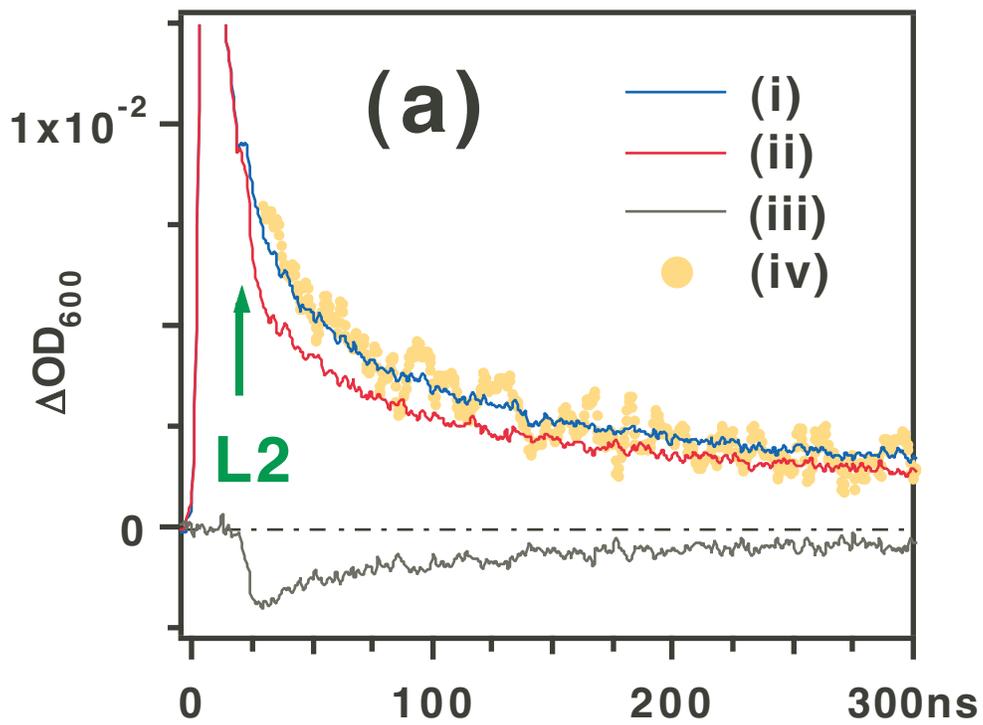

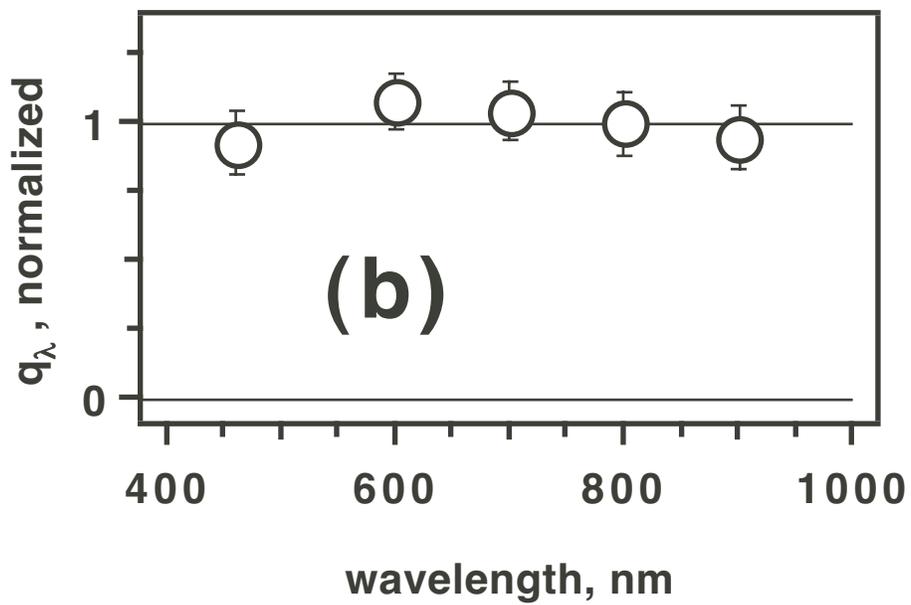

Fig. 10; Shkrob et al.

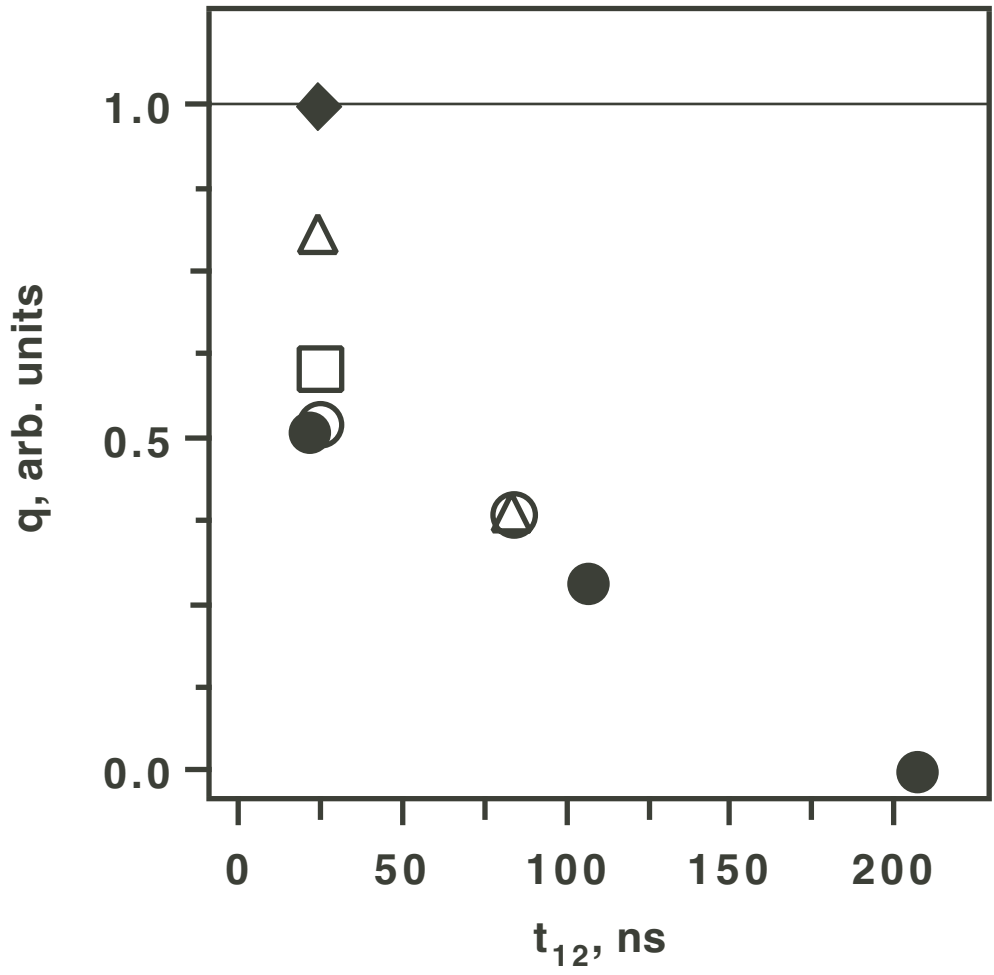

**Fig. 11; Shkrob et al.**

**SUPPLEMENTARY MATERIAL**                         **jp0000000**

## Supporting Information.

**Captions to figures 1S to 7S.**

**Fig. 1S**

A family of normalized TA kinetics in the visible obtained in 355 nm laser photolysis of oxygenated aqueous solution of $TiO_2$ nanoparticles (similar excitation conditions to Fig. 1). These kinetics were normalized at the maximum delay time of $t = 300$ ns. Note the double logarithmic plot. The wavelengths of analyzing light are given in the color scale.

**Fig. 2S**

Same as Fig. 1S, for oxygen-free solution. The TA kinetics in the near IR (out to $t = 8$ μm) were obtained using a relatively slow Ge detector (GMP-566), so that only the data points taken after the first 20 ns are shown. The use of this detector was preferable to the use of a faster InGaAs detectors because the response time was wavelength-independent (see ref. [] for more detail).

**Fig. 3S**

The progression of TA spectra in the near IR. These spectra were obtained using a fast Ge photodiode under the same excitation conditions as Fig. 3b from a 5 vol % glycerol solution of aqueous $TiO_2$ nanoparticles. The integration windows are given in the legend.

**Fig. 4S**

A family of normalized TA kinetics (450 to 950 nm) for 5 vol % glycerol solution (Figs. 3b and 3S). These kinetics are normalized at ca. 200 ns, by which delay time the hole scavenging by glycerol is complete. The difference between these normalized TA kinetics in the visible and at $\lambda \geq 900$ nm is due to the additional absorbance signal from short-lived holes (on top of the signal from the electrons); this extra TA signal systematically increases towards the blue. Only trapped electrons absorb in the near IR.

**Fig. 5S**

Functions $f_\lambda(t)$ defined by eq. (3) for (a) $\lambda = 1.1$ μm and (b) $\lambda = 1.3$ μm vs. glycerol concentration (in vol %, indicated in the color scale). Same excitation conditions as in Fig. 6. These TA kinetics were obtained using a fast InGaAs photodiode. (c) Fractions $\Phi_e$ obtained at 900 nm *(open circles),* 1100 nm *(open squares),* and 1300 nm *(open diamonds)* are shown. All such plots for $\lambda \geq 900$ nm are identical within the experimental error. The solid line is a least squares fit to eq. (2). These observations suggest that the absorbance of the hole in the near IR is negligible, and the TA signal originates from electrons only.

**Fig. 6S**

Same as Fig. 10, for 1064 nm photoexcitation (6 ns FWHM, 91 mJ). The high fluence of 1064 nm photons is needed due to the low quantum yield of the photobleaching, ca. 0.024 (section 3.2). For 532 nm photoexcitation, this yield is ca. 12 times higher.

**Fig. 7S**

The data of Fig. 2 plotted as a function of the photon energy. The solid lines are Gaussian curves. The indexing is the same except for trace (i) which corresponds to trace (iv) in Fig. 2.

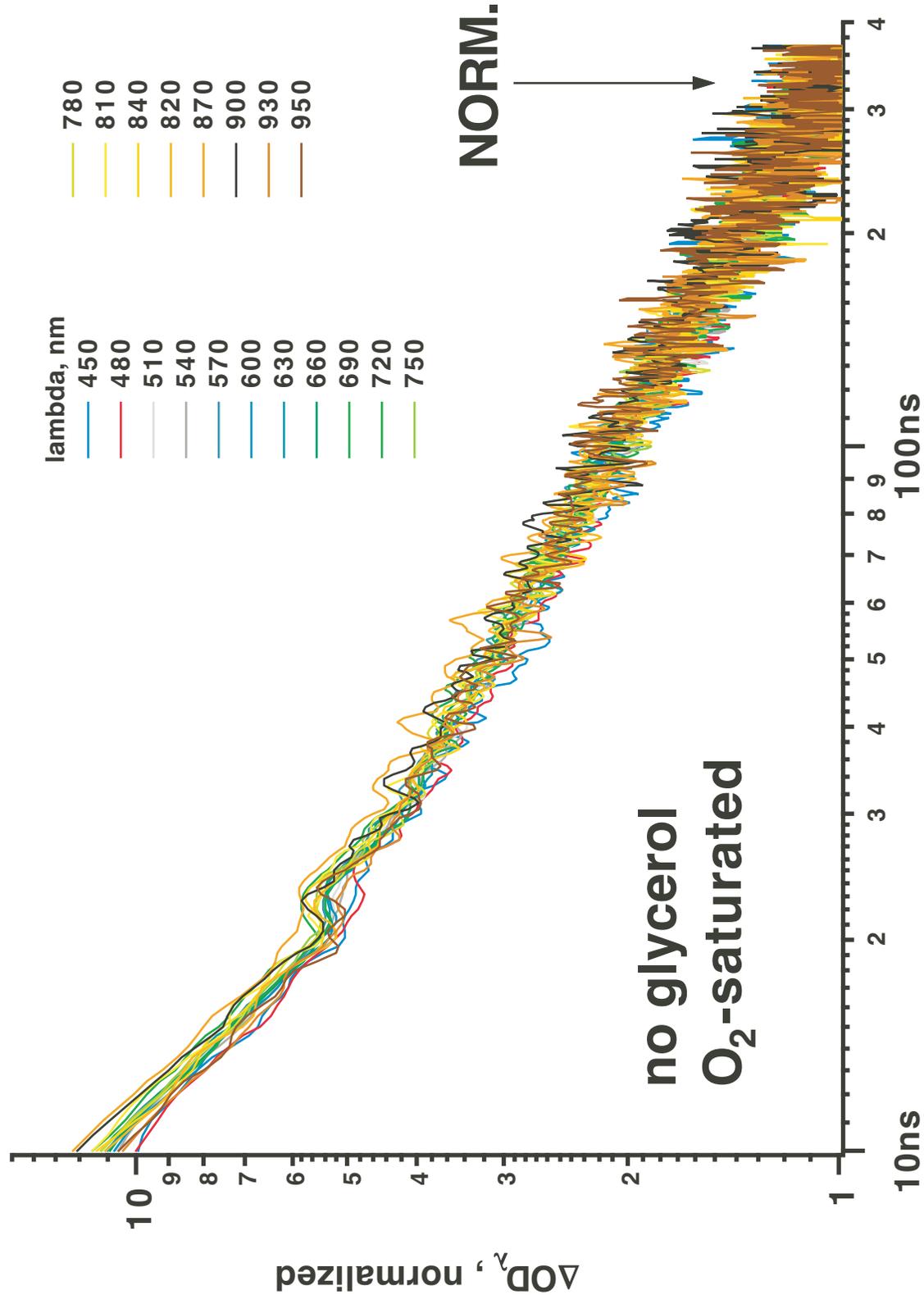

Fig. 1S; Shkrob et al.

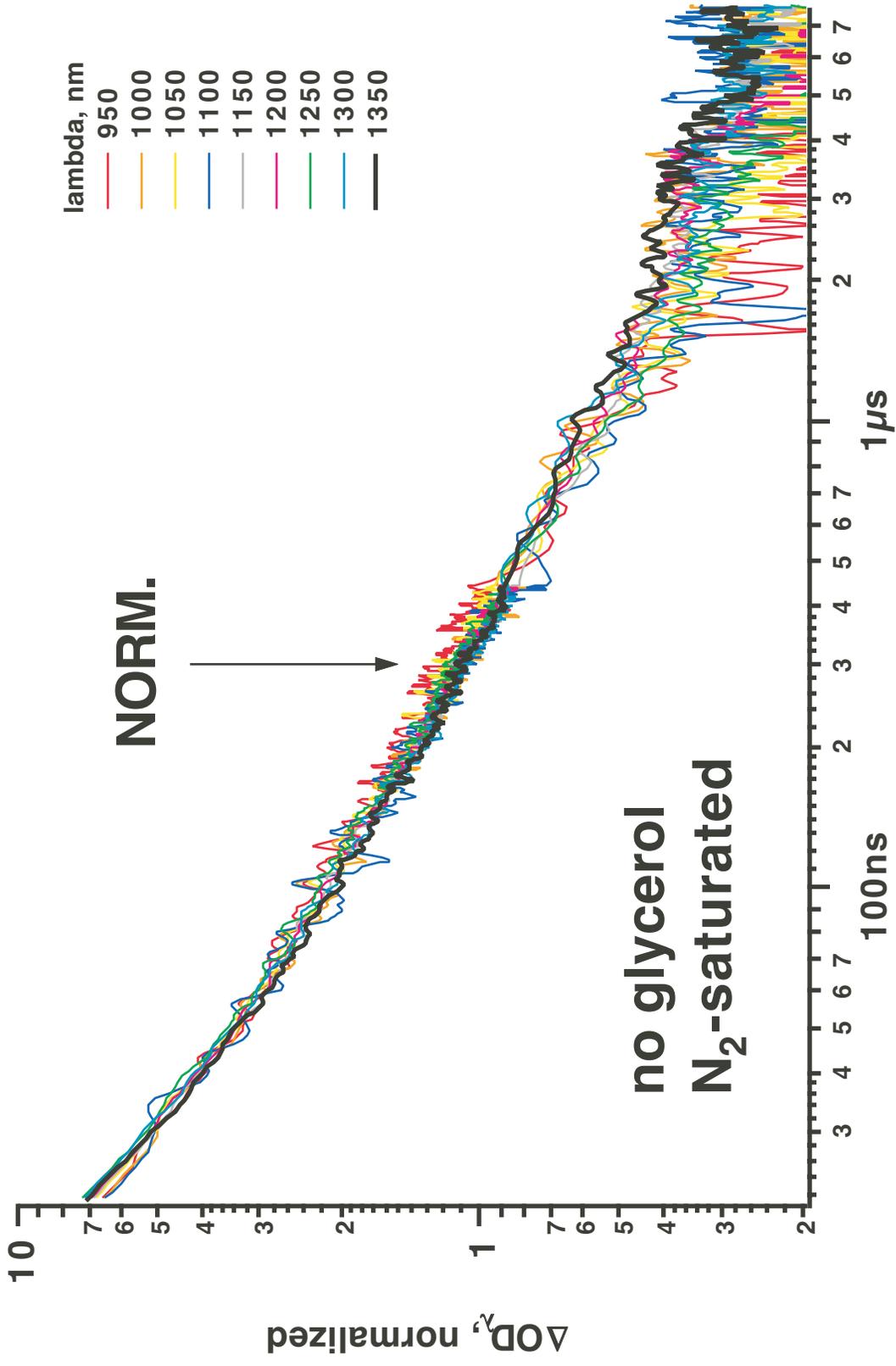

**Fig. 2S; Shkrob et al.**

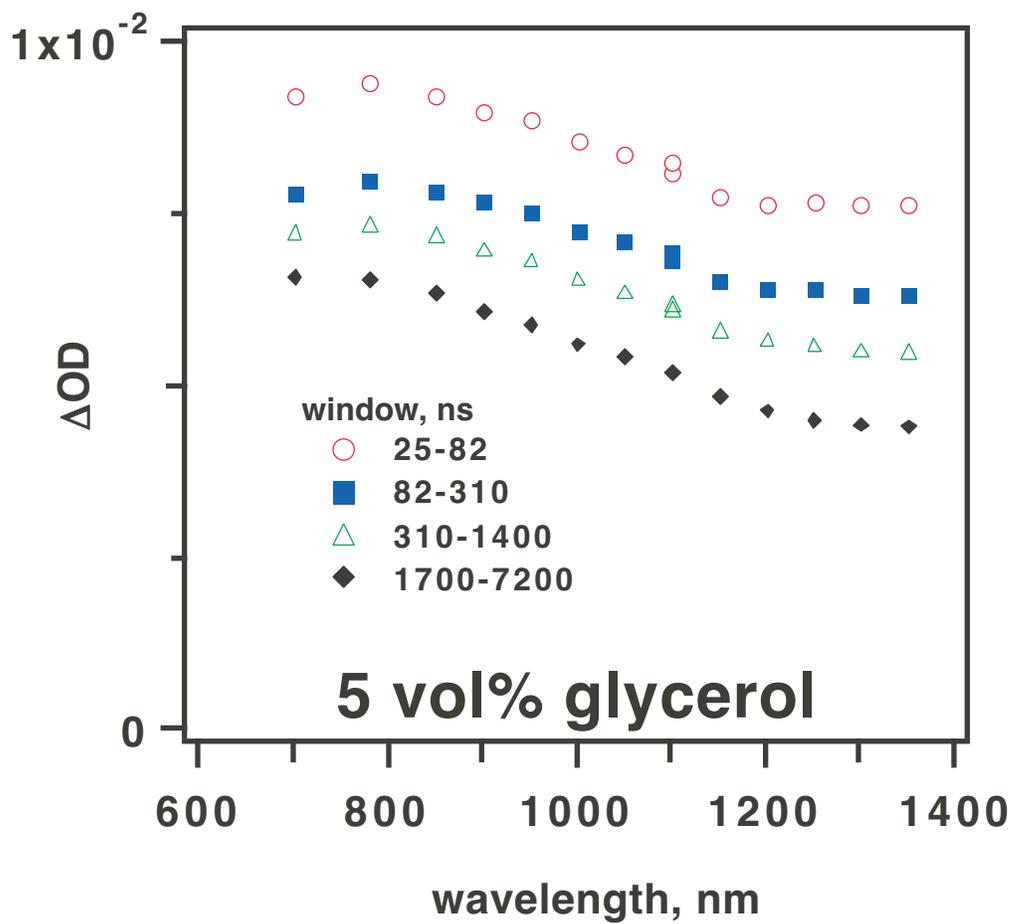

Fig. 3S; Shkrob et al.

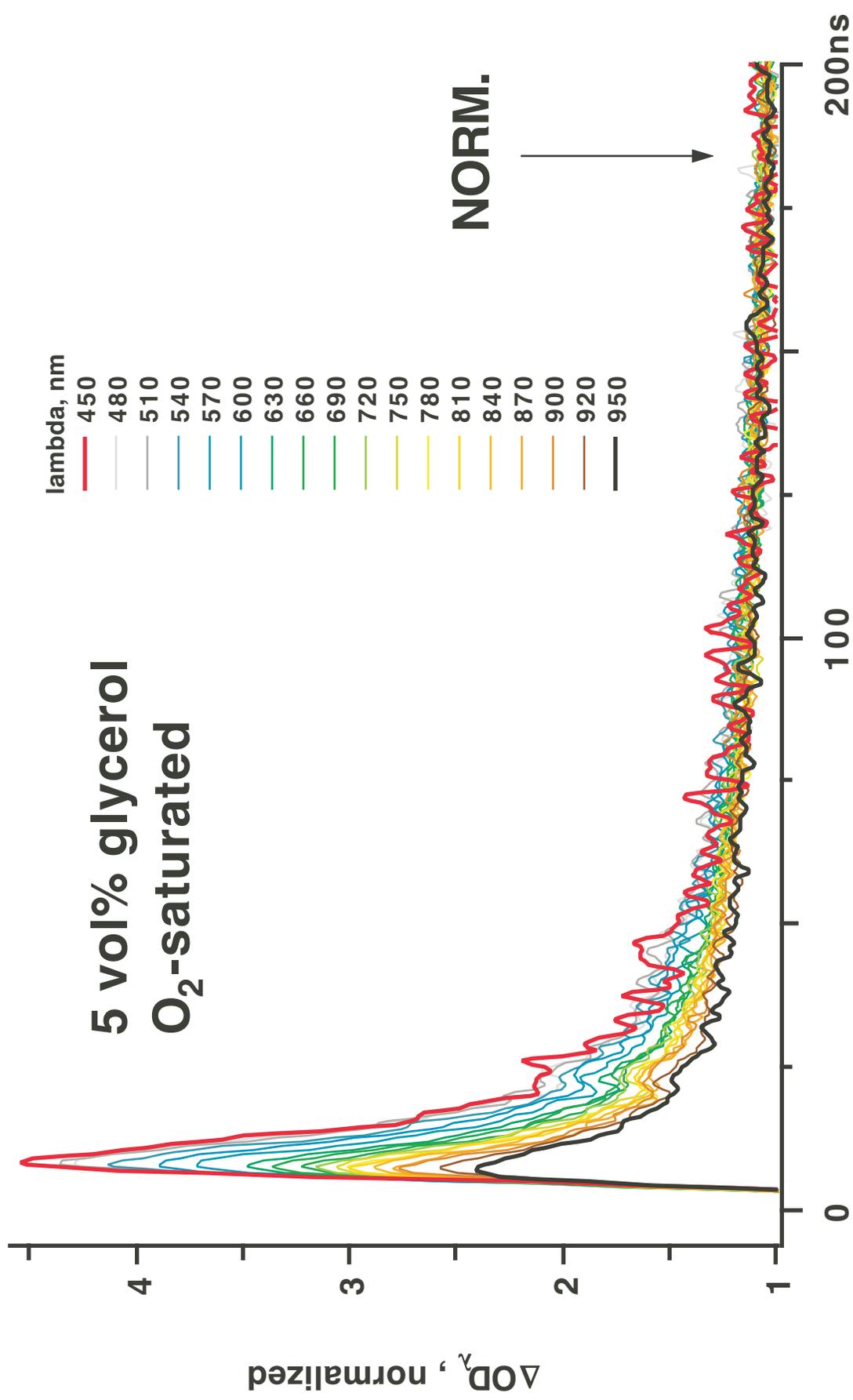

Fig. 4S; Shkrob et al.

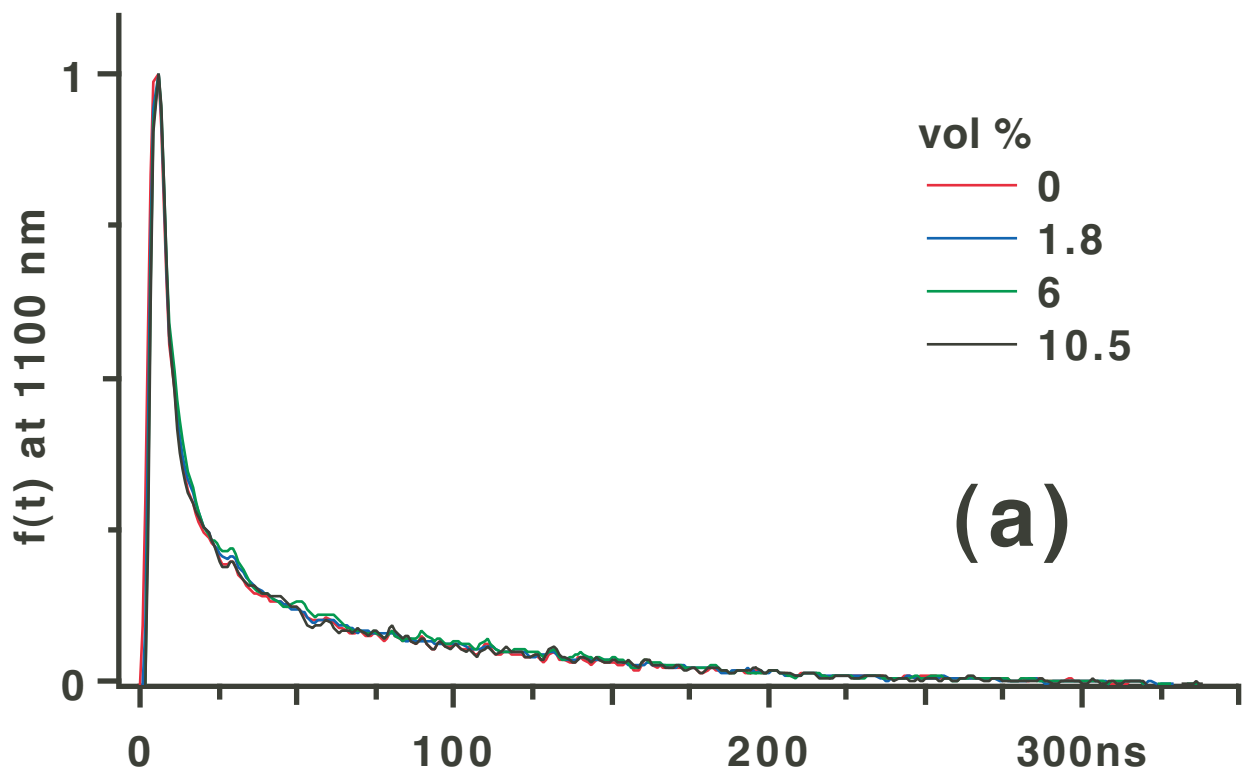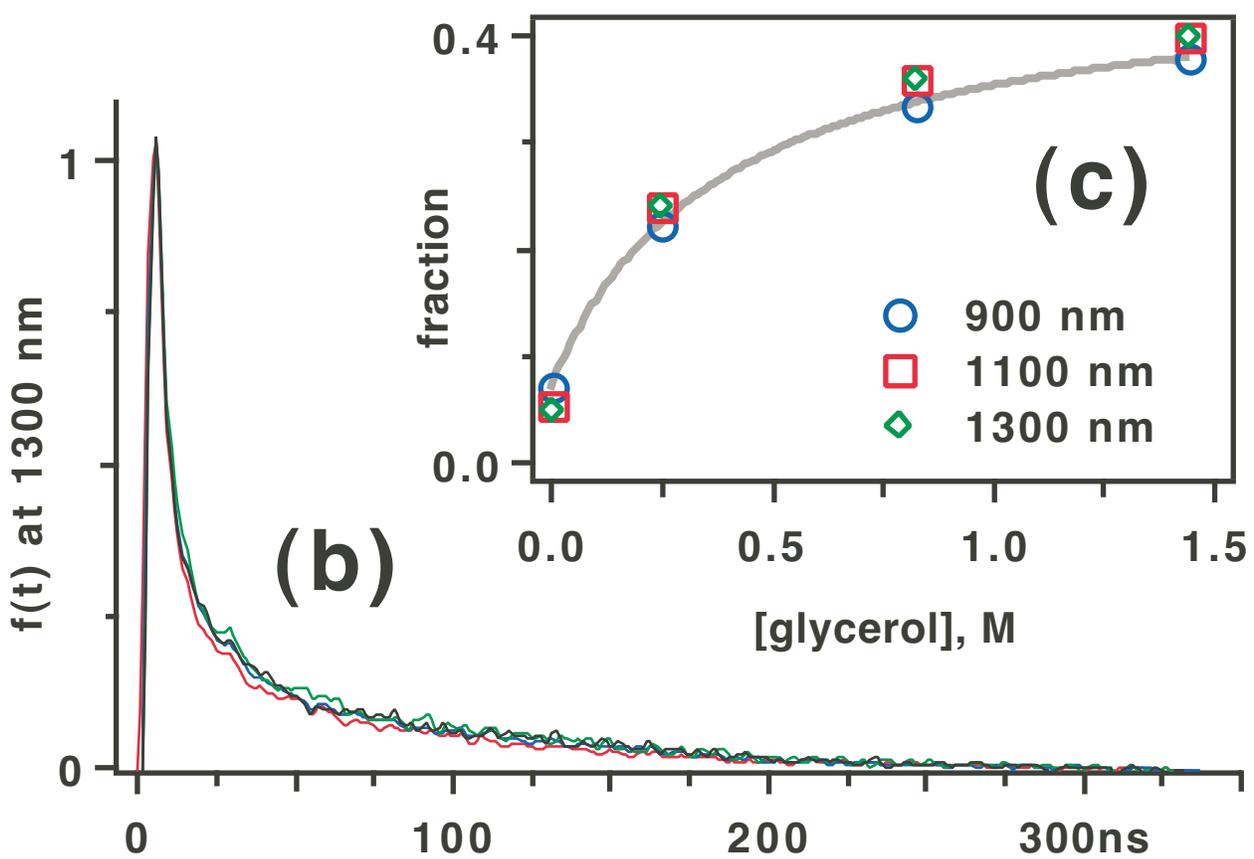

Fig. 5S; Shkrob et al.

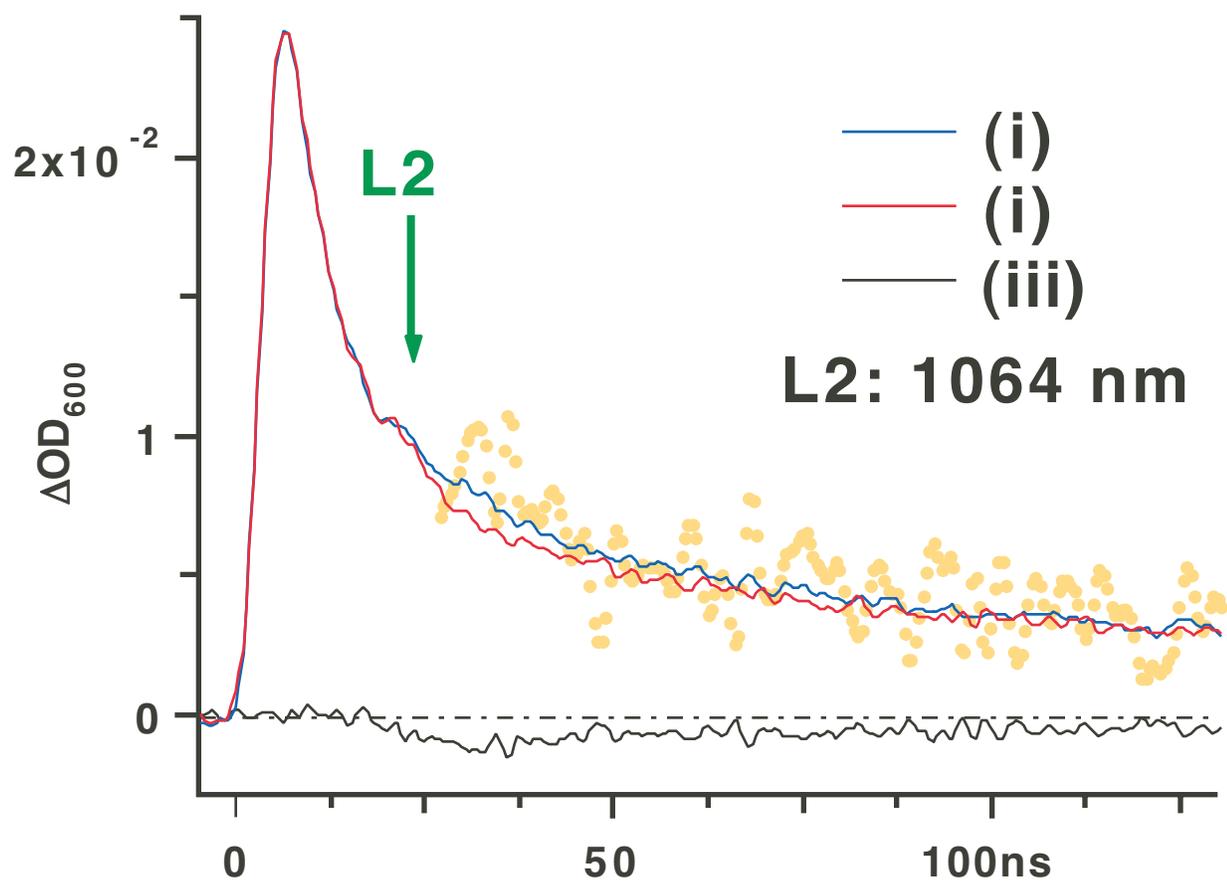

Fig. 6S; Shkrob et al.

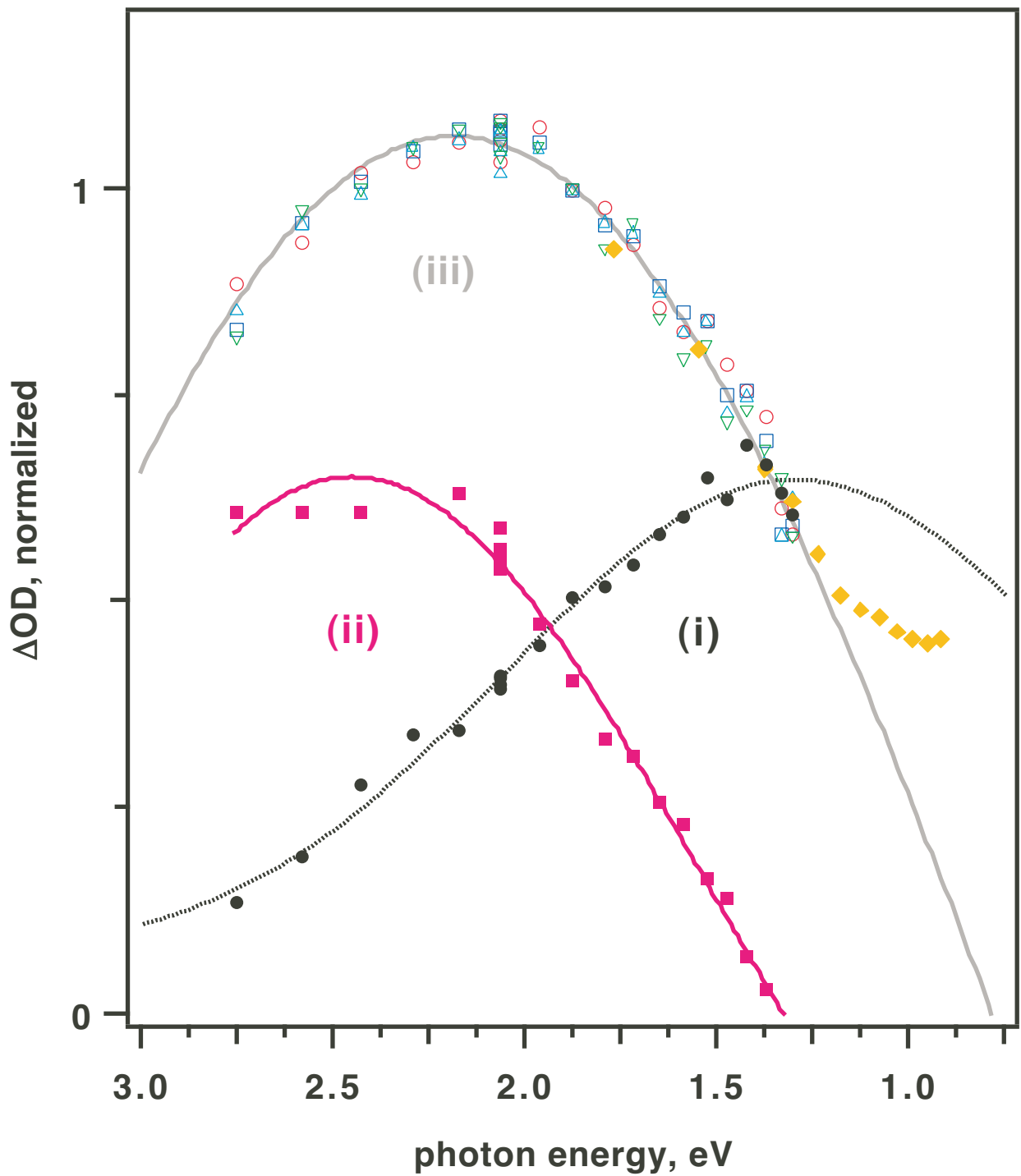

Fig. 7S; Shkrob et al.